\documentstyle[epsf]{mn}

\newif\ifAMStwofonts
\AMStwofontstrue

\newcommand{\odn}[1]{(\ref{#1})}
\newcommand{\ts}{\tau_{\rm T}}
\newcommand{\msol}{M$_\odot$}

\def\g{$\gamma$}

\ifoldfss
  \ifCUPmtlplainloaded \else
    \NewTextAlphabet{textbfit} {cmbxti10} {}
    \NewTextAlphabet{textbfss} {cmssbx10} {}
    \NewMathAlphabet{mathbfit} {cmbxti10} {} 
    \NewMathAlphabet{mathbfss} {cmssbx10} {} 
  \fi
  \ifAMStwofonts
    \ifCUPmtlplainloaded \else
      \NewSymbolFont{upmath} {eurm10}
      \NewSymbolFont{AMSa} {msam10}
      \NewMathSymbol{\upi}     {0}{upmath}{19}
      \NewMathSymbol{\umu}     {0}{upmath}{16}
      \NewMathSymbol{\upartial}{0}{upmath}{40}
      \NewMathSymbol{\leqslant}{3}{AMSa}{36}
      \NewMathSymbol{\geqslant}{3}{AMSa}{3E}

      \let\leq=\leqslant \let\le=\leqslant
      \let\geq=\geqslant 
    \fi
  \fi
\fi 

\ifnfssone
  \newmathalphabet{\mathit}
  \addtoversion{normal}{\mathit}{cmr}{m}{it}
  \addtoversion{bold}{\mathit}{cmr}{bx}{it}
  \newmathalphabet{\mathbfit} 
  \addtoversion{normal}{\mathbfit}{cmr}{bx}{it}
  \addtoversion{bold}{\mathbfit}{cmr}{bx}{it}
  \newmathalphabet{\mathbfss} 
  \addtoversion{normal}{\mathbfss}{cmss}{bx}{n}
  \addtoversion{bold}{\mathbfss}{cmss}{bx}{n}
  \ifAMStwofonts
    \ifCUPmtlplainloaded \else
      \UseAMStwoboldmath
      \makeatletter
      \new@mathgroup\upmath@group
      \define@mathgroup\mv@normal\upmath@group{eur}{m}{n}
      \define@mathgroup\mv@bold\upmath@group{eur}{b}{n}
      \edef\UPM{\hexnumber\upmath@group}
      \new@mathgroup\amsa@group
      \define@mathgroup\mv@normal\amsa@group{msa}{m}{n}
      \define@mathgroup\mv@bold\amsa@group{msa}{m}{n}
      \edef\AMSa{\hexnumber\amsa@group}
      \makeatother
      \mathchardef\upi="0\UPM19
      \mathchardef\umu="0\UPM16
      \mathchardef\upartial="0\UPM40
      \mathchardef\leqslant="3\AMSa36
      \mathchardef\geqslant="3\AMSa3E

      \let\leq=\leqslant \let\le=\leqslant
      \let\geq=\geqslant 
    \fi
  \fi
\fi 

\ifnfsstwo
  \DeclareMathAlphabet{\mathbfit}{OT1}{cmr}{bx}{it}
  \SetMathAlphabet\mathbfit{bold}{OT1}{cmr}{bx}{it}
  \DeclareMathAlphabet{\mathbfss}{OT1}{cmss}{bx}{n}
  \SetMathAlphabet\mathbfss{bold}{OT1}{cmss}{bx}{n}
  \ifAMStwofonts
    \ifCUPmtlplainloaded \else
      \DeclareSymbolFont{UPM}{U}{eur}{m}{n}
      \SetSymbolFont{UPM}{bold}{U}{eur}{b}{n}
      \DeclareSymbolFont{AMSa}{U}{msa}{m}{n}
      \DeclareMathSymbol{\upi}{0}{UPM}{"19}
      \DeclareMathSymbol{\umu}{0}{UPM}{"16}
      \DeclareMathSymbol{\upartial}{0}{UPM}{"40}
      \DeclareMathSymbol{\leqslant}{3}{AMSa}{"36}
      \DeclareMathSymbol{\geqslant}{3}{AMSa}{"3E}

      \let\leq=\leqslant \let\le=\leqslant
      \let\geq=\geqslant 
    \fi
  \fi
\fi 

\ifCUPmtlplainloaded \else
  \ifAMStwofonts \else 
    \def\upi{\pi}
    \def\umu{\mu}
    \def\upartial{\partial}
  \fi
\fi

\title[Thermal synchrotron radiation and its Comptonization]
{Thermal synchrotron radiation and its Comptonization\\
in compact X-ray sources}

\author[G. Wardzi\'nski and A. A. Zdziarski]
{{Grzegorz Wardzi\'nski$^*$ and Andrzej A. Zdziarski\thanks{E-mail:
gwar@camk.edu.pl, aaz@camk.edu.pl}}\\
N. Copernicus Astronomical Center, Bartycka 18, 00-716 Warsaw, Poland \\
}

\date{Accepted 1999 November 23. Received 1999 June 14}

\pagerange{\pageref{firstpage}--\pageref{lastpage}}
\pubyear{1999}

\begin{document}

\maketitle

\label{firstpage}

\begin{abstract}
We investigate the process of synchrotron radiation from thermal electrons at
semi-relativistic and relativistic temperatures.  We find an analytic
expression for the emission coefficient for random magnetic fields with
an accuracy significantly higher than of those derived previously. We
also present analytic approximations to the synchrotron turnover frequency,
treat Comptonization of self-absorbed synchrotron radiation, and give simple
expressions for the spectral shape and the emitted power. We also consider
modifications of the above results by bremsstrahlung.

We then study the importance of Comptonization of thermal synchrotron radiation
in compact X-ray sources.  We first consider emission from hot accretion flows
and active coronae above optically-thick accretion discs in black-hole binaries
and AGNs.  We find that for plausible values of the magnetic field strength,
this radiative process is negligible in luminous sources, except for those with
hardest X-ray spectra and stellar masses.  Increasing the black-hole mass
results in a further reduction of the maximum Eddington ratio due to this
process.  Then, X-ray spectra of intermediate-luminosity sources, e.g.,
low-luminosity AGNs, can be explained by synchrotron Comptonization only if
they come from hot accretion flows, and X-ray spectra of very weak sources are
always dominated by bremsstrahlung. On the other hand, synchrotron
Comptonization can account for power-law X-ray spectra observed in the low
states of sources around weakly-magnetized neutron stars.
\end{abstract}

\section{Introduction}

The theory of cyclotron and synchrotron radiation is a well established part
of physics.  However, there still remain uncertainties about the accuracy and
the range of applicability of some analytic formulae describing the emission.
One important example of such uncertainties concerns the spectra of synchrotron
emission from mildly-relativistic and relativistic thermal plasma, in which
case numerous studies devoted to this field, e.g., Jones \& Hardee (1979),
Petrosian (1981, hereafter P81), Takahara \& Tsuruta (1982), Petrosian \&
McTiernan (1983), Robinson \& Melrose (1984), Mahadevan, Narayan \& Yi (1996,
hereafter MNY96), yielded results not entirely consistent with each other.

Precise determination of  spectra from the thermal synchrotron process is of
key significance for studies of emission from accretion flows onto black holes
and neutron stars. Although direct, optically-thin, thermal synchrotron
emission is rarely observable, accurate optically-thin spectra are necessary to
determine the turnover frequency, below which the plasma becomes
optically-thick. The resulting partially-absorbed spectrum is then observable
in some cases. Furthermore, photons from that spectrum provide a supply of seed
photons for Comptonization in the plasma, which process gives rise to
observable power-law spectra with high-energy cutoffs. On the other hand,
blackbody radiation emitted by optically-thick accretion discs and other forms
of optically-thick matter may constitute a competing supply of seed photons. In
addition, bremsstrahlung radiation is always emitted by a hot plasma.

In this work, we first consider formulae for the synchrotron emission
coefficient of a thermal plasma (Section \ref{s:synrad}).  We clarify the
accuracy and the range of applicability of previously derived formulae as well
as propose our own expressions.  We concentrate on mildly relativistic and
relativistic plasmas, as exhaustive studies of non-relativistic thermal
cyclotron emission exist (Chanmugam et al.\ 1989 and references therein).  We
then calculate the turnover frequency and consider effects of bremsstrahlung
emission and self-absorption.  Second, we investigate Comptonization of the
synchrotron radiation (hereafter abbreviated as the CS process) and present
convenient formulae for the resulting spectra and luminosities (Section
\ref{s:compt}).  In Section \ref{applications}, we apply our results to two
main geometries of accretion flows onto a black hole:  a hot, two-temperature,
optically thin disc, and active regions above a cold disc (i.e., a patchy
corona).  Those two models, for various plausible prescriptions for the
magnetic field strength, are then compared with spectral data from black-hole
binaries and AGNs.  Finally, in Section \ref{neutron}, we study the importance
of the CS proces for formation of spectra of accreting neutron stars with
weak/moderate magnetic fields.

\section{Synchrotron radiation from thermal plasmas}
\label{s:synrad}

\subsection{Emission of a single electron}

Let us consider an electron (with charge $e$) moving in a uniform magnetic
field, $\bmath{B}$, at a velocity, $\bmath{\beta} \equiv \bmath{v} /c$.  Let
$\xi$ be the angle between $\bmath{v}$ and $\bmath{B}$, and $\vartheta$ be the
angle between $\bmath{B}$ and the direction towards the observer.  Then, the
cyclo-synchrotron power per unit frequency and unit solid angle in the observer
frame and in cgs units is given by (e.g.\ Bekefi 1966; Pacholczyk 1970)
\begin{eqnarray}
\label{pojed}
\eta_\nu \equiv \lefteqn{ \frac{{\rm d} W}{{\rm d} \nu \, {\rm d} \Omega \,
{\rm d}t} (\vartheta, \xi, \gamma) = \frac{2 \upi e^2 \nu^2}{c} \times }\\
\lefteqn{\sum_{n=1}^{\infty} \delta (y_n) \left[ \left(\frac{\cos\vartheta -
\beta \cos\xi }{\sin\vartheta}\right)^2 J^2_n (z) + \beta^2 \sin^2 \xi J'^2_n
(z)
\right],}\nonumber
\end{eqnarray}
where
\begin{equation} y_n \equiv \frac{n \nu_{\rm c}}{\gamma} - \nu (1-\beta \cos\xi
\cos\vartheta), \,\,  z \equiv \frac{\nu \gamma \beta \sin\vartheta
\sin \xi }{\nu_{\rm c}}, \end{equation}
$\nu_{\rm c} \equiv e B/2 \upi m_{\rm e} c$ is the cyclotron frequency, $\gamma
= (1-\beta^2)^{-1/2}$ is the Lorentz factor, $J_n$ is a Bessel function of
order $n$, and $m_{\rm e}$ is the electron mass.

Note that an additional factor of $(1- \beta \mu \cos\vartheta)^{-1}$ due to
the Doppler effect should have appeared in the formal derivation of equation
\odn{pojed}. However, this term disappears in the case of an electron moving
chaotically, as, e.g., in a thermal plasma. For detailed discussion, see
Scheuer (1968), Rybicki \& Lightman (1979, section 6.7), Pacholczyk (1970,
section 3) and references in these works.

\subsection{Synchrotron emission coefficient in a thermal plasma}

The emission coefficient, $j_\nu(\vartheta)$, of a thermal plasma at a
temperature, $T$, and with a uniform magnetic field, $\bmath{B}$, can be
obtained by integrating the rate of equation \odn{pojed} over a relativistic
Maxwellian electron distribution,
\begin{equation} \label{max} n_{\rm e} (\gamma) = \frac{n_{\rm e}}{
\Theta} {\gamma (\gamma^2-1)^{1/2} \over K_2(1/\Theta)} \exp \left(- {\gamma
\over \Theta}\right), \end{equation}
where $\Theta\equiv kT/m_{\rm e} c^2$ is the dimensionless plasma temperature,
and $n_{\rm e}$  is the electron density. $K_2$ is a modified Bessel
function, which can be approximated by
\begin{eqnarray} \label{k2}
\lefteqn{ K_2\left(1\over \Theta\right) \approx} \\
\lefteqn { \cases{
\left(\upi \Theta\over 2\right)^{1/2} \left(1+ {15 \Theta\over 8} +
{105\Theta^2\over
128} -0.203 \Theta^3 \right) {\rm e}^{-1/\Theta}, & $\Theta\leq 0.65$;\cr
2\Theta^2-{1\over 2} +{\ln(2\Theta)+3/4-\gamma_{\rm E}\over 8\Theta^2}
+ {\ln(2\Theta)+0.95\over 96\Theta^4}, & $\Theta\geq 0.65$,\cr} \nonumber }
\end{eqnarray}
where $\gamma_{\rm E}\approx 0.5772$ is Euler's constant and the last
coefficients in the series have been adjusted to achieve a relative error $\leq
0.0008$.

The integral for the emission coefficient is then,
\begin{equation} \label{jotnu} j_\nu(\vartheta) = \int_1^\infty {\rm d}
\gamma\, {1 \over 2} n_{\rm e} (\gamma) \int_{-1}^{1} {\rm d} \mu\,
\eta_\nu(\vartheta, \mu, \gamma) ,
\end{equation}
where $\mu = \cos\xi$. This integration is relatively difficult to carry out
due to the complicated form of the integrand and the presence of $J_n$ in
$\eta_\nu$. In particular, standard numerical methods of computing $J_n$ become
very inefficient for $n\gg 1$. P81 obtained an approximate solution by
replacing the summation over $n$ by integration over $\nu$ in equation
\odn{pojed} and then using the principal term of an asymptotic expansion of
$J_n(z)$, where $0\leq z\leq n \beta<n$ [see equation (9.3.7) of Abramowitz \&
Stegun (1970, hereafter AS70)]. The region of validity of that approximation of
$J_n$ is given by $\gamma^2 \ll \nu /\nu_{\rm c}$. The integration over $\mu$
and $\gamma$ by the method of the steepest descent with some additional
approximations yields
\begin{equation}
\label{jasy}
j_\nu (\vartheta) = \frac{2^{1/2} \upi e^2
n_{\rm e} \nu}{3 c \, K_2(1/\Theta)} \exp \left[ - \left( {9 v \over
2\sin\vartheta } \right)^{1/3} \right].
\end{equation}
where $v\equiv \nu/\nu_{\rm c}\Theta^2$.
We see that the emission coefficient becomes very small both at large
frequencies, and at viewing angles, $\vartheta$, significantly different from
$\upi/2$.

We note that equation (26) in P81, which corresponds to the formula \odn{jasy},
is too small by a factor of 2 due to the adopted normalization of the electron
distribution [equation (23) in P81] being also twice too small.  On the other
hand, Takahara \& Tsuruta (1982) obtain a formula [equation (2.9) in their
paper] for the absorption coefficient for the perpendicular polarization of
synchrotron radiation, $\alpha_\nu^\perp$, which agrees with our expression
(\ref{jasy}) for the emission coefficient at $\vartheta=\upi/2$.  (They also
compute the coefficient for the parallel polarization, $\alpha_\nu^\parallel$,
which, however, is negligibly small compared to the dominant coefficient of the
perpendicular polarization.)  We note that the relation between the absorption
and emission coefficients is given by Kirchhoff's law, which, in the case of
polarized radiation, contains the source function for each polarization
separately, i.e., $B^\parallel_\nu = B^\perp_\nu = B_\nu/2$ (see, e.g.\
Chanmugam et al.\ 1989).  The coefficients for polarized radiation are then
related to the coefficients including both polarizations by $j_\nu=
j^\parallel_\nu + j^\perp_\nu$, $\alpha_\nu = ({\alpha}^{\parallel}_{\nu}
+{\alpha}^{\perp}_{\nu})/2$.  We also note that the emission coefficient
computed by Jones \& Hardee (1979) for $\Theta \gg 1$ is larger than the
corresponding limit of equation (\ref{jasy}) by a factor of $2^{1/2}$, which
discrepancy we have not been able to explain.

In our numerical calculations, we used an approximation of $J_n(z)$ in terms of
the Airy function [equation (9.3.6) in AS70], which maximum relative error is
$\sim 0.08$ at $n=1$, $z=0$, but it rapidly decreases for larger values of $n$,
$z$.  If the argument of the Airy function is $<2.25$, we approximate it by its
power series [equation (10.4.3) in AS70] up to the 13th power.  Otherwise, we
use the approximation to $J_n$ of P81 described above, in which case the
maximum error is $\sim 0.02$.  Typical relative accuracy of the resulting
approximation to $J_n$ is then $< 0.01$.  The accuracy can be further increased
by adding the first-order correction to the principal term of the asymptotic
expansion of $J_n$, see equation (9.3.7) in AS70.  The derivative, $J'_n$, is
calculated with the second expression in equation (9.1.30) of AS70. Note that
the above method is more accurate, but also more complicated, than a related
method given by Wind \& Hill (1971).

\begin{figure}
\begin{center}
\leavevmode
\epsfxsize=7.2cm
\epsfbox{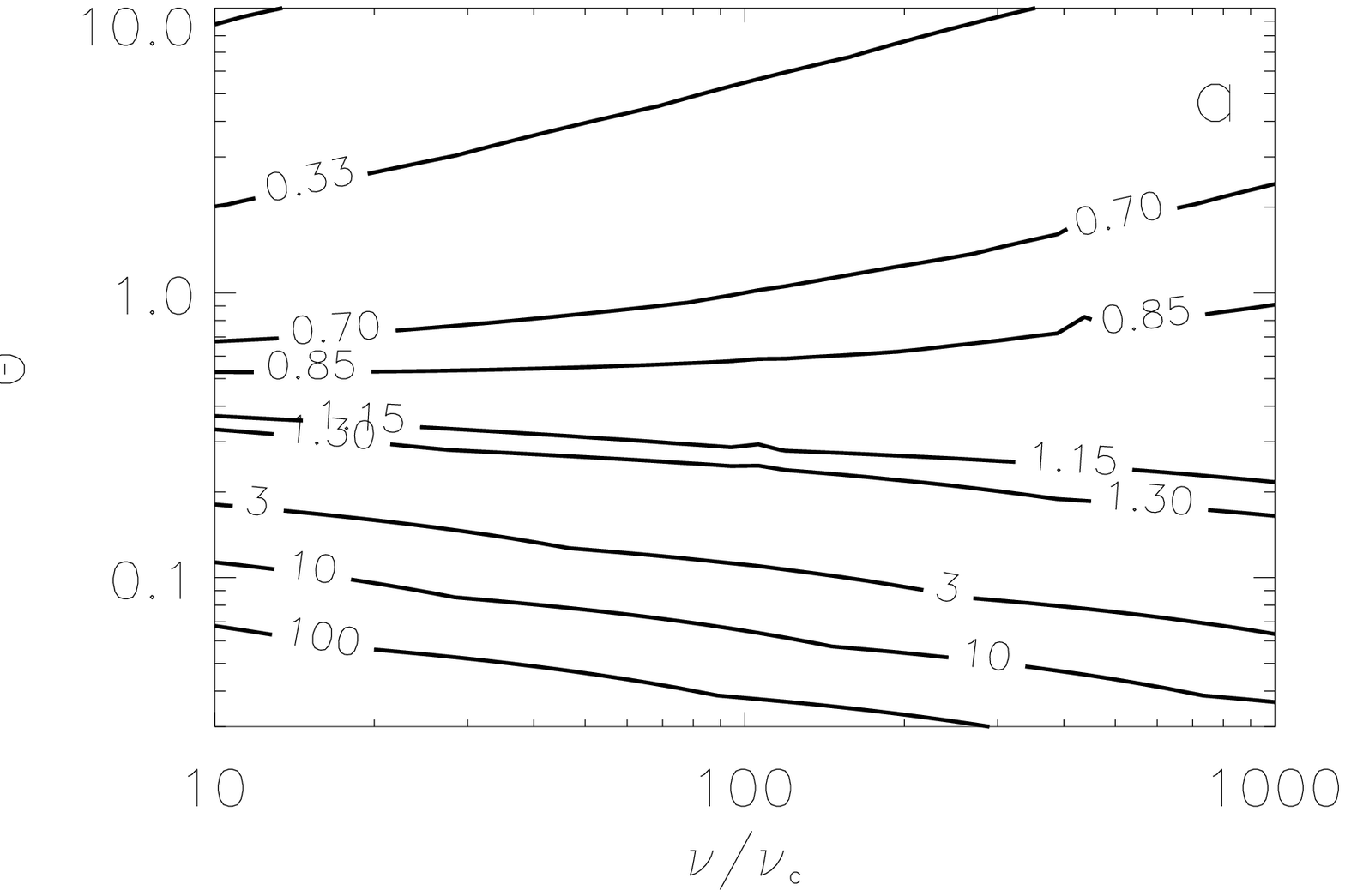}
\end{center}
\nobreak
\begin{center}
\leavevmode
\epsfxsize=7.2cm
\epsfbox{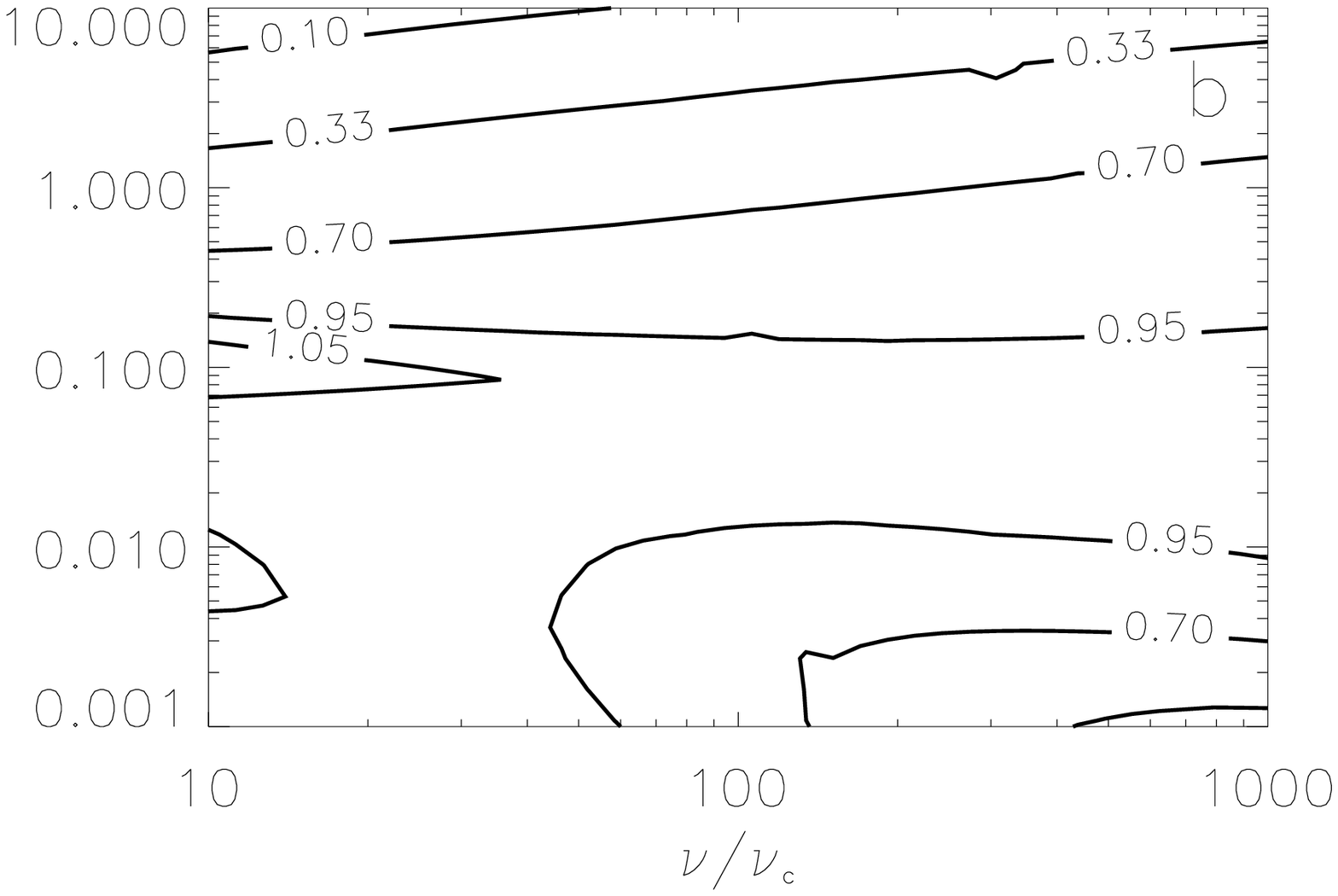}
\end{center}
\caption{(a) Contour plots of the ratio of the approximated emission
coefficient
of equation (\protect\ref{jasy}) to the exact numerical value (at
$\vartheta=\upi/2$). (b) The same for equations
(\protect\ref{jotth})-(\protect\ref{gz}).}
\label{fangle}
\end{figure}

We have tested the accuracy of formula \odn{jasy} compared to the results of
numerical integration (which themselves are in a very good agreement with
numerical results tabulated for $\Theta\la 0.1$ by Chanmugam et al.\ 1989) of
equation \odn{jotnu} at $\vartheta=\upi/2$ for $10 \le \nu / \nu_{\rm c} \le
1000$ and $10^{-3} \le \Theta \le 10$. These ranges include parameters
most relevant for compact sources (e.g.\ Takahara \& Tsuruta 1982; Zdziarski
1986, hereafter Z86; Narayan \& Yi 1995, hereafter NY95), and we hereafter use
them while testing other synchrotron rates. Figure \ref{fangle}a shows the
results for $\Theta>0.02$. We see that equation \odn{jasy} overestimates the
actual value of $j_\nu$ by orders of magnitude at low values of $\Theta$, and
should not be used at all for $\Theta \la 0.1$. The relative accuracy strongly
depends on $\Theta$ (being best at $\Theta \sim 0.4$) and it improves with
increasing $\nu/\nu_{\rm c}$. In the relativistic limit, $\Theta \gg 1$, the
accuracy of formula \odn{jasy} depends on $v$ only.

We thus see that the accuracy of the emission coefficient (\ref{jasy}) is not
satisfactory for detailed modelling of astrophysical plasmas, as also found by
MNY96. We suggest using a significantly more accurate expression [eq.\ (11) of P81 with
additional approximations of eq.\ (25) of P81 and eq.\ (20) of Petrosian \&
McTiernan (1983)],
\begin{eqnarray}
\label{jotth}
\lefteqn{ j_\nu(\vartheta) = \frac{\upi
e^2}{2\,c} \left(\nu \nu_{\rm c}\right)^{1/2} {\cal X}(\gamma_0) n_{\rm e}
(\gamma_0) \left(1+2 {\cot^2\vartheta \over \gamma_0^2}\right)} \nonumber \\
\lefteqn{\qquad \times \left[1- (1-\gamma_0^{-2}) \cos^2\vartheta\right]^{1/4}
{\cal Z}(\vartheta,\gamma_0). }
\end{eqnarray}
Here
\begin{eqnarray}
\label{defz}
\lefteqn{ {\cal Z}(\vartheta,\gamma) =\left[ \frac{t
\exp{\left[(1+t^2)^{-{1\over
2}}\right]}}{1+ (1+t^2)^{1\over 2}}\right]^{2n}\!,\,\, t\equiv
(\gamma^2-1)^{1\over 2} \sin \vartheta, } \\
\label{dx}
\lefteqn{ n \equiv \frac{\nu (1+t^2)}{\nu_{\rm c} \gamma}, \quad
{\cal X}(\gamma) =\cases{\left[\frac{2 \Theta
\left(\gamma^2-1\right)}{\gamma\left(3\gamma^2-1\right)}\right]^{1/2}, &
$\Theta \la 0.08$;
\cr
\left(\frac{2 \Theta}{3 \gamma}\right)^{1/2}, & $\Theta \ga 0.08$, }}
\end{eqnarray}
and
\begin{equation}
\label{gz}
\gamma_0=\cases{\left[1+\left(\frac{2\,\nu\Theta}{\nu_{\rm
c}}\right)\left(1+\frac{9\, \nu \Theta \sin^2\vartheta}{2\, \nu_{\rm
c}}\right)^{-{1\over 3}}\right]^{1\over 2}\! , & $\Theta \la 0.08$;\cr
\left[1+\left(\frac{4\, \nu \Theta}{3\, \nu_{\rm c} \sin\vartheta
}\right)^{2\over 3}\right]^{1\over 2}, & $\Theta \ga 0.08$,}
\end{equation}
where $\gamma_0$ is the Lorentz factor of the saddle point of the integral
(\ref{jotnu}) over $\gamma$ (i.e.\ of those  thermal electrons that
contribute most to the emission at $\nu$), corresponding to the minimum
of $n_{\rm e}(\gamma) {\cal Z}(\vartheta,\gamma) /\gamma$. Note that
approximations
\odn{dx} and \odn{gz} have a small discontinuity at $\Theta \simeq 0.08$.

Figure \ref{fangle}b shows the relative accuracy of the approximation of
equations \odn{jotth}-\odn{gz}. We see it is much more accurate than expression
\odn{jasy} throughout the test range, with the minimum relative error
at $0.01 \la \Theta \la 0.2$. Also, the relative accuracy of this approximation
varies with $\Theta$ and $\nu/\nu_{\rm c}$ much slower than that of equation
\odn{jasy}. At $\Theta \sim 10^{-3}$ the approximation \odn{jotth}-\odn{gz}
matches well the nonrelativistic emission coefficient given by equations
(14)-(15) of Trubnikov (1958).

Our expression (\ref{jotth})-(\ref{gz}) can be compared with that of Robinson
\& Melrose (1984), who have also used a method based on that of P81 with
accuracy improved with respect to that paper.  We have found that their
expression, in the form of Dulk (1985) and  with the non-relativistic
Maxwellian replaced by the relativistic one, is slightly more accurate, but
also more complicated, than expression (\ref{jotth})-(\ref{gz}). We also note
that a relativistic generalization of the results of Robinson \& Melrose (1984)
by Hartmann, Woosley \& Arons (1988) appears incorrect. Namely, their
expression (B1) should be multiplied by a factor $2\upi (\gamma^2_0 -
1)^{-1/2}$ (in their notation).

\subsection{Angle-averaged emission coefficient}
\label{s:average}

The emission coefficients derived above are appropriate for a plasma in a
uniform magnetic field. However, in typical astrophysical conditions, we expect
either to deal with emission from regions where magnetic fields are chaotic or
to observe radiation being a sum of contributions from a number of regions with
different orientations of magnetic field. Therefore, we would like to find the
emission coefficient averaged over magnetic field directions (or, equivalently,
over the viewing angle), $\bar{j}_\nu= \epsilon_\nu/4 \upi$, where
$\epsilon_\nu$ is the emitted power per unit volume and frequency, and
\begin{equation} \label{srednia}
\bar{j}_\nu  = {1 \over {4 \upi}} \int j_\nu (\vartheta)  {\rm d}\Omega = {1
\over 2} \int j_\nu (\vartheta) \sin\vartheta \; {\rm d}\vartheta.
\end{equation}

To obtain an expression for $\bar j_\nu$, we integrate equation \odn{jotth}
over $\vartheta$ using the method of the steepest descent with the saddle point
at $\vartheta = \upi /2$. Here, we treat ${\cal Z}$ as
the fast varying (with $\vartheta$) part of the integrand \odn{jotth}.
The resulting expression is,
\begin{equation}
\label{calk}
\bar{j}_\nu = \frac{\upi^{3\over 2}e^2\left(\nu\nu_{\rm
c}\right)^{1\over2}n_{\rm e}(\gamma_0){\cal Z}(\vartheta,\gamma_0) {\cal
X}(\gamma_0)}{
2^{3\over 2}c} \left|  \partial^2 \ln {\cal Z}(\vartheta,\gamma_0) \over
\partial \vartheta^2 \right|^{-{1\over2}}
\end{equation}
to be evaluated at $\vartheta=\upi/2$.

On the other hand, the asymptotic emission coefficient (\ref{jasy}) can be
integrated over $\vartheta$, as done by MNY96. This yields,
\begin{equation}
\label{mahadevan}
\bar{j}_\nu = {2^{1/6} \upi^{3/2} e^2 n_{\rm e} \nu \over 3^{5/6} c K_2 (1
/\Theta) v^{1/6} } a(\Theta,v)
\exp \left[ - \left( 9 v \over 2\right)^{1/3} \right],
\end{equation}
where the correction factor, $a$,
represents the ratio of the exact emission coefficient to that obtained by
integration of equation (\ref{jasy}). This factor can be calculated from the
ratio of equations (33) to (31) in MNY96 with fitting coefficients of their
Table 1.

\begin{figure}\begin{center}\leavevmode
\epsfxsize=7.2cm \epsfbox{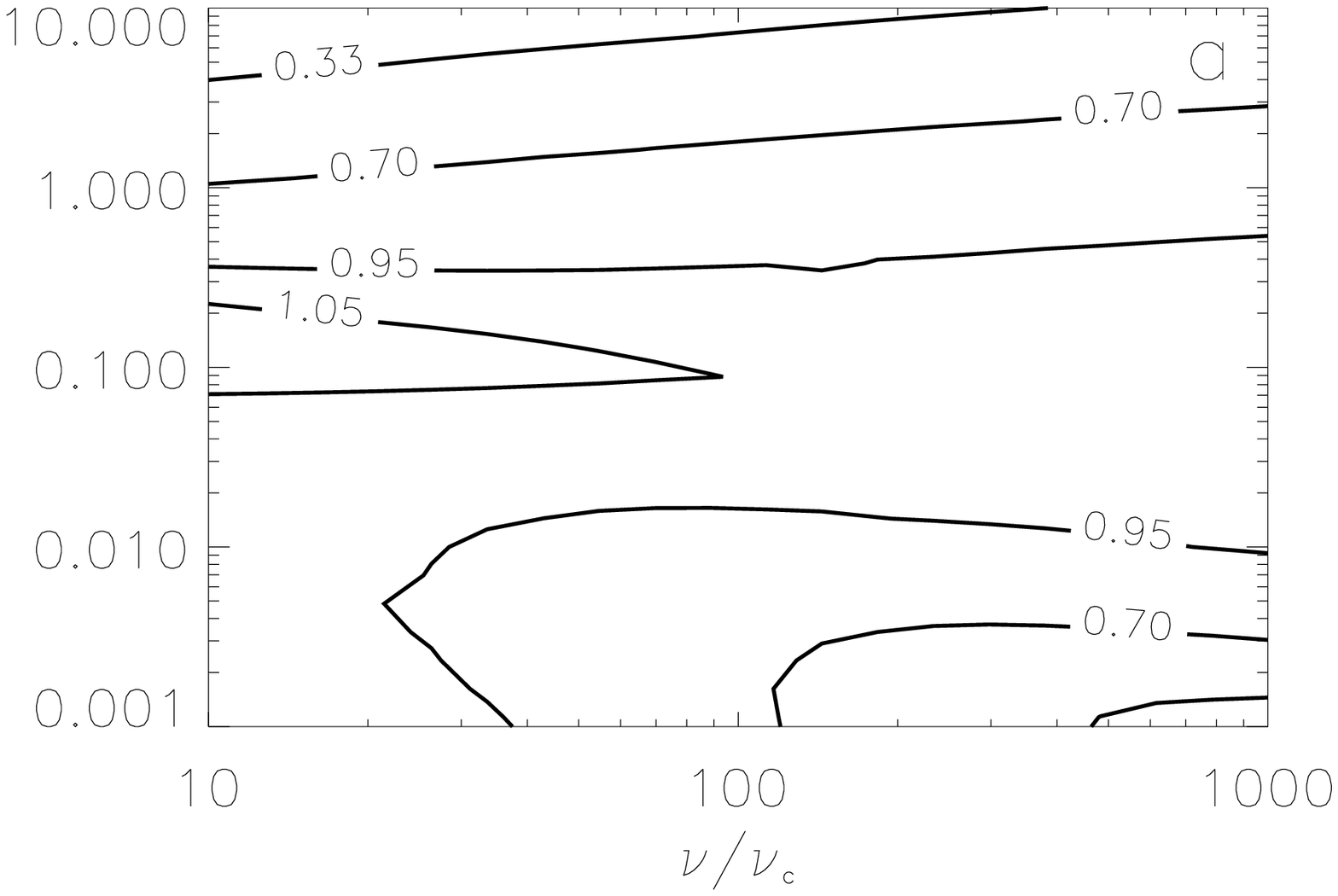}\end{center}\nobreak
\begin{center}\leavevmode
\epsfxsize=7.2cm \epsfbox{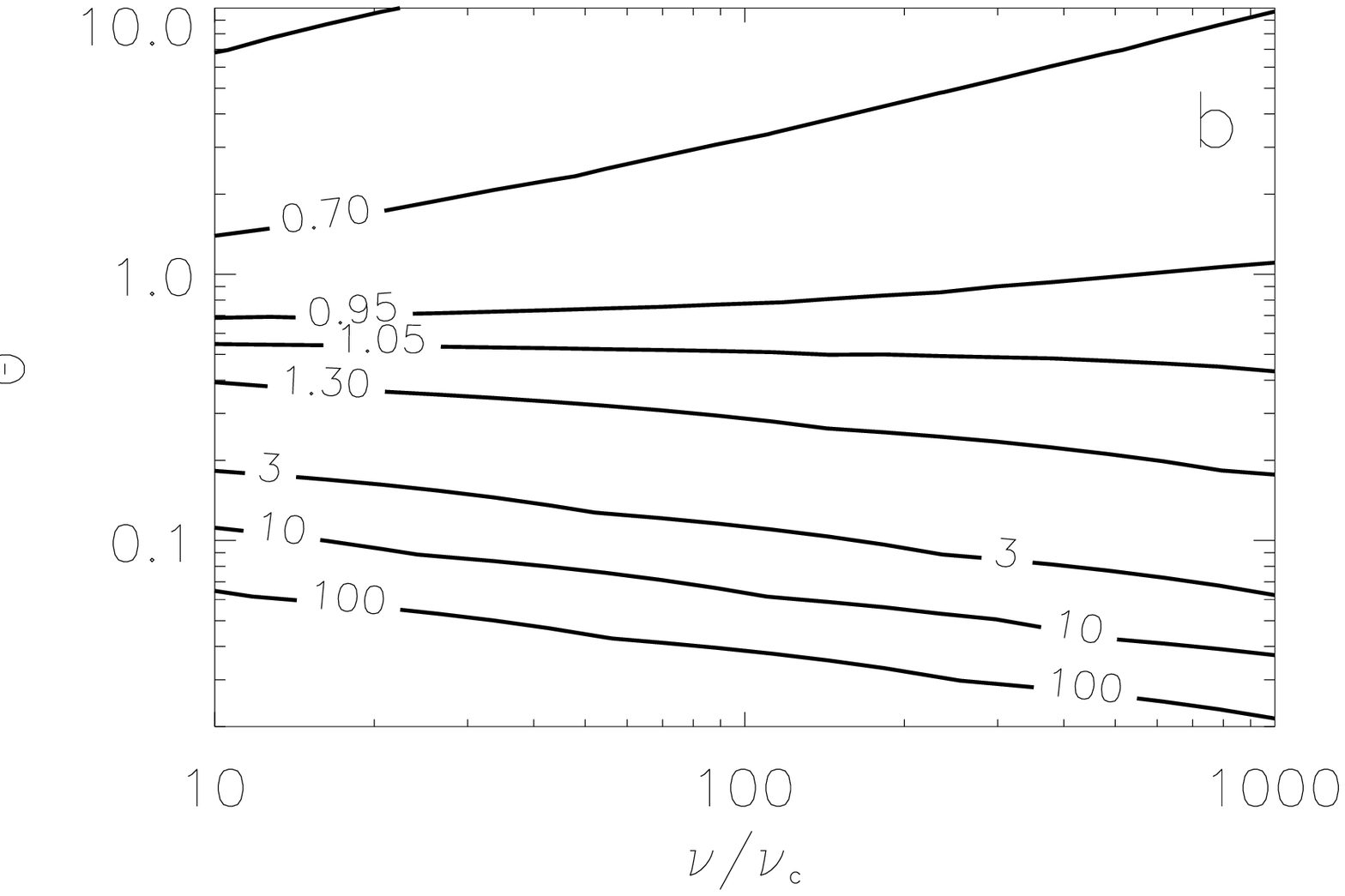}\end{center}
\caption{(a) Contour plots of the ratio of the approximated emission
coefficient
\odn{calk} to the correct numerical value. (b) The same for equation
\odn{mahadevan} with $a\equiv 1$.
}\label{f:cdm}
\end{figure}

We have tested the accuracy of the formulae \odn{calk} and \odn{mahadevan}, the
latter both with and without the tabulated corrections, again in the range
$10^{-3} \leq \Theta \leq 10$ and $10 \le \nu/\nu_{\rm c} \le 1000$. Figure
\ref{f:cdm}a shows the relative error of equation \odn{calk}. We see that the
relative error is typically $\la 30$ per cent except for $\Theta > 1$ with the
best accuracy at $0.02 \la \Theta \la 0.4$.

The expression \odn{mahadevan} with the tabulated corrections of MNY96 matches
very well the numerical results (the difference is less than a few per cent),
with the exception of the case of $\Theta=0.084$, where the corrections in
Table 1 of MNY96 appear misprinted as the obtained values are a few times too
small and the error increases with frequency.

The accuracy of formula \odn{mahadevan} with $a\equiv 1$ is shown in Figure
\ref{f:cdm}b. We see that it provides an estimate correct within a factor of
$\sim 3$ or better only for $\Theta > 0.1$ and the best accuracy is obtained
at $\Theta \sim 0.6$ This formula strongly overestimates the correct result for
$\Theta \la 0.1$. At relativistic temperatures, $\Theta \gg 1$, the relative
accuracy of equation (\ref{mahadevan}) with $a\equiv 1$ depends on $v$ only. It
underestimates somewhat the actual emission coefficient, and its integration
over $\nu$ yields a value of the total emitted power too low by a factor of
$3^{1/2} 35\upi/2^8\simeq 0.744$ with respect to the actual power,
\begin{equation}\label{e_total}
\epsilon = 16\Theta^2 n_{\rm e} \sigma_{\rm T} c {B^2\over 8\upi}, \quad \Theta
\gg 1, \end{equation}
where $\sigma_{\rm T}$ is the Thomson cross section. Then, equation
(\ref{mahadevan}) with $a\equiv 1$ and divided by 0.744 represents an
approximation to the ultra-relativistic thermal synchrotron emission
coefficient satisfying the total-power constraint (\ref{e_total}).

\subsection{Synchrotron self-absorption}
\label{ss:turn}

The synchrotron radiation is self-absorbed by electrons up to the turnover
frequency, $\nu_{\rm t}$, above which the plasma becomes optically thin to the
synchrotron radiation, i.e.,
\begin{equation}
\label{defturn}
\tau = \alpha_{\nu_{\rm_t}} R =1,
\end{equation}
where $R$ is the characteristic size of the plasma. Below $\nu_{\rm t}$, the
observed spectrum has the blackbody form, typically in the Rayleigh-Jeans
limit.
In this limit and averaging over angles, Kirchhoff's law implies,
\begin{equation}
\label{thin}
{\bar{j}_{\nu_{\rm t}} \over 2 \nu_{\rm t}^2 m_{\rm e} \Theta} R = 1.
\end{equation}
This can be solved numerically. On the other hand, Zdziarski et al.\ (1998)
give the solution using equation \odn{mahadevan},
\begin{equation}
\label{nut}
{\nu_{\rm t} \over \Theta^2 \nu_{\rm c}} = \frac{343}{36} \ln^3 {C\over
\ln{C\over \ln{C\over \dots}}}, \,\,\,
C=  \frac{3}{7 \Theta} \left[\upi \tau_{\rm T} a\exp(1/\Theta)\over
3  \alpha_{\rm f} x_{\rm c} \right]^{2\over 7}\! ,
\end{equation}
where $x_{\rm c}\equiv h\nu_{\rm c}/m_{\rm e} c^2$ is a dimensionless cyclotron
frequency, $\tau_{\rm T} =n_{\rm e}\sigma_{\rm T} R$ is the Thomson optical
depth of the plasma, and $\alpha_{\rm f}$ is the fine-structure constant. Note
that $\nu_{\rm t}$ depends on $n_{\rm e}$ and $R$ only through their product,
or equivalently, $\ts$. Typically, the correction factor, $a$, is  a
slowly-varying function of frequency (MNY96) and then equation (\ref{nut}) for
$\nu_{\rm t}$ is explicit. We also see that the dependence of $\nu_{\rm t}$ on
$a$ is only logarithmic, and thus the accuracy of determining the synchrotron
emission coefficient only weakly affects the turnover frequency.

Mahadevan (1997) used a slightly different method of determining $\nu_{\rm t}$.
Namely, he equated the total flux of the Rayleigh-Jeans emission to that in the
optically-thin synchrotron one. This corresponds to setting the right-hand side
of equation (\ref{thin}) to 3/4 instead of 1, which changes slightly the
resulting value of $\nu_{\rm t}$ as compared to equation (\ref{nut}).

\begin{figure}\begin{center}\leavevmode
\epsfxsize=7.2cm \epsfbox{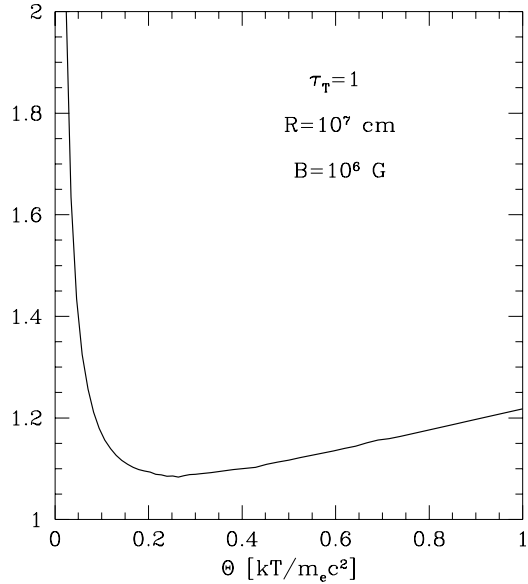}\end{center}
\caption{The ratio of the turnover frequency calculated from equation
\odn{nut} with $a\equiv 1$ to the accurate numerical result as a function of
temperature.
}\label{f:trnaccur}
\end{figure}

We have compared the approximated formula \odn{nut} with $a\equiv 1$ with
results of numerical calculations. It turns out that the error of equation
\odn{nut} is almost independent of the values of the parameters other then
$\Theta$. Figure \ref{f:trnaccur} compares the results of the two methods in
the temperature range $0.03 \leq \Theta \leq1$ for a source with $\ts=1$,
$R=10^7\,$cm and $B=10^6\,$G. (In this parameter range, the effect of
bremsstrahlung can be neglected, see below.) We see that for $0.1 \la \Theta
\la 1$ the discrepancy does not exceed 20 per cent but grows rapidly for lower
temperatures. We have also found a power-law approximation for $x_{\rm t}\equiv
h\nu_{\rm t}/m_{\rm e} c^2$ as
\begin{equation}\label{nu_ratio}
x_{\rm t} \approx 2.6\times 10^{-6} \left(\Theta\over 0.2 \right)^{0.95}
\ts^{0.05} \left(B\over 10^6\,{\rm G}\right)^{0.91},
\end{equation}
which is accurate to $\la 30$ per cent for $0.05\la \Theta \la 0.4$, $0.3\la
\ts\la 3$, $10\,{\rm G} \la B \la 10^8$ G. Typical values of $x_{\rm t}/x_{\rm
c}$ at mildly-relativistic temperatures are in the range of 10--$10^3$.

\subsection{Effect of bremsstrahlung on the turnover frequency}
\label{turnover}

We note that the above formulae for the turnover frequency do not take into
account bremsstrahlung emission and self-absorption.  However, for some
combinations of $\Theta$, $n_{\rm e}$ and $B$, bremsstrahlung emission can be
comparable to or stronger than the synchrotron one and then the turnover
frequency determined neglecting bremsstrahlung will be incorrect. Then, the
absorption coefficient, $\alpha_{\nu_{\rm t}}$, should include a contribution
from bremsstrahlung, $\alpha^{\rm ff}_{\nu}$. Note that since $\alpha^{\rm
ff}_{\nu}\propto n_{\rm e}^2 R$, $\nu_{\rm t}$ is no more solely a function of
$\ts$ whenever bremsstrahlung is important.

The effect of bremsstrahlung on the turnover frequency can be checked by
computing $\nu_{\rm t}$ without taking into account bremsstrahlung, and then
computing the emission coefficient, $\bar{j}_{\nu_{\rm_t}}$, including both
processes. As long as $\bar{j}_{\nu_{\rm_t}} \gg j^{\rm ff}_{\nu_{\rm_t}}$,
equations in Section \ref{ss:turn} can be used. We hereafter use formulae for
the free-free emission coefficient of Svensson (1984).

\begin{figure}
\begin{center}\noindent\epsfxsize=7.2cm \epsfbox{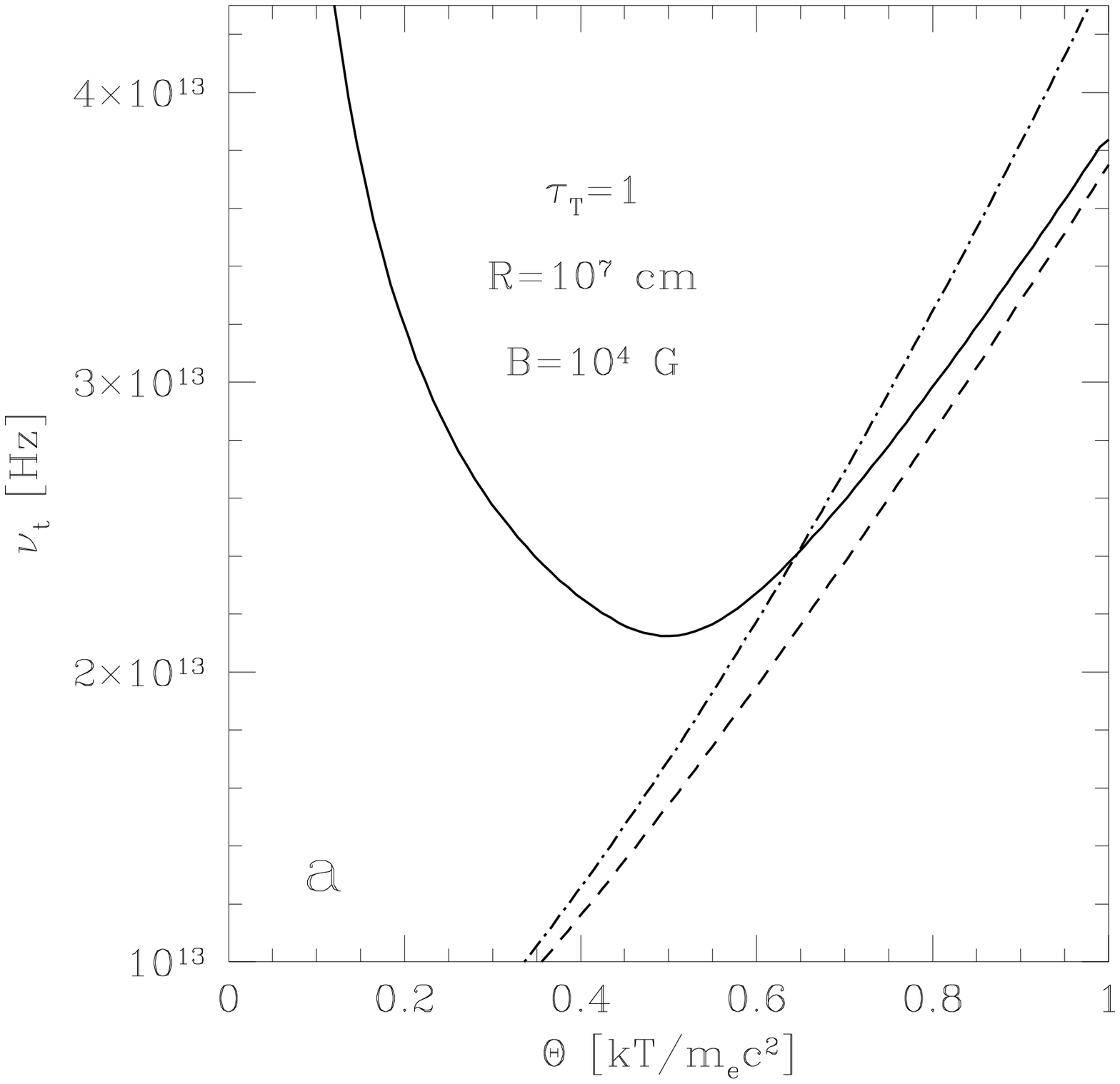}\end{center}
\nobreak
\begin{center}\noindent\epsfxsize=7.2cm \epsfbox{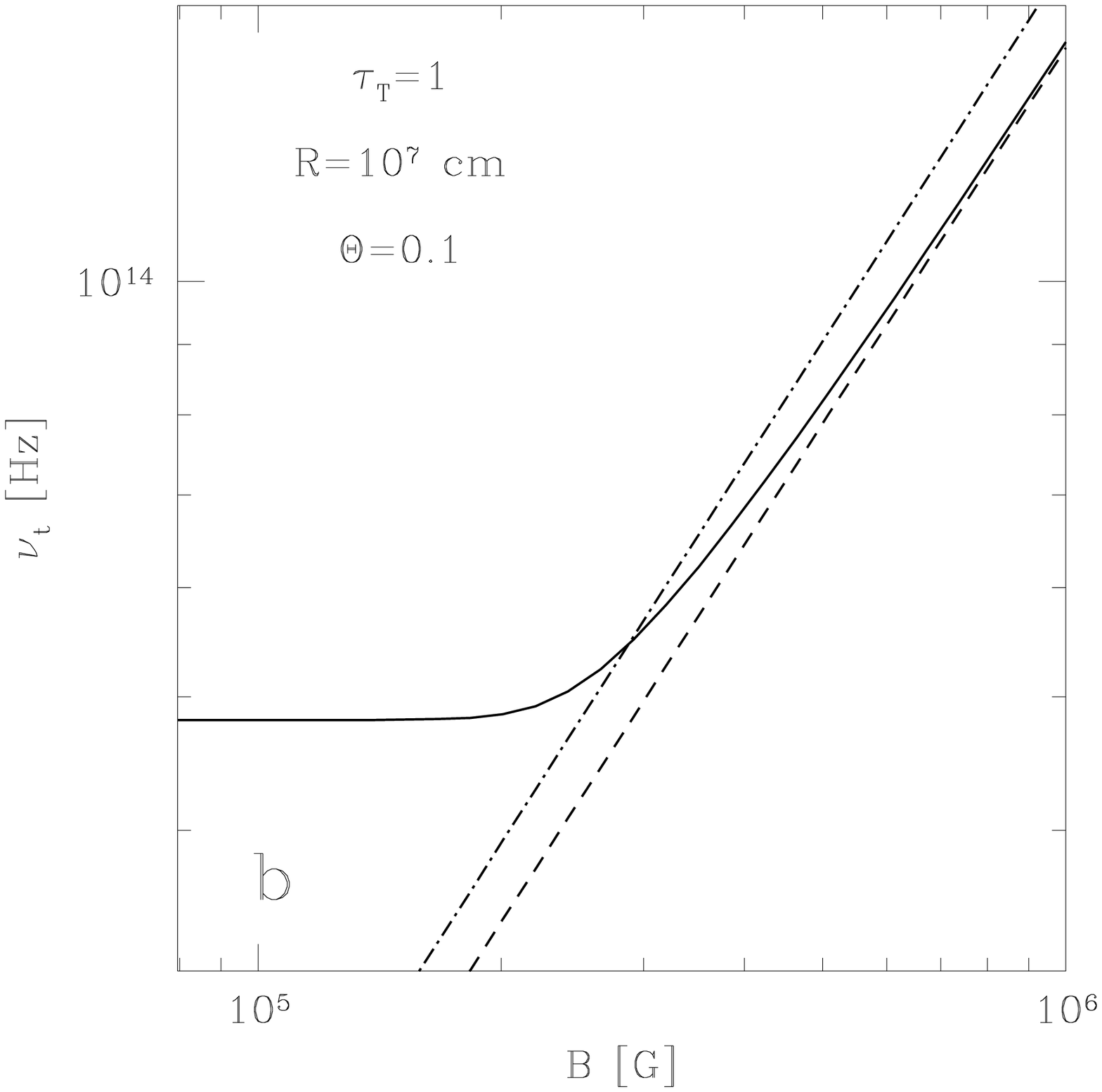}\end{center}
\caption{Example dependences of the turnover frequency on (a) temperature and
(b) magnetic field strength. The dot-dashed and dashed curves give $\nu_{\rm
t}$ calculated with the analytic approximation of equation \odn{nut}, and
numerically using equation \odn{thin}, respectively, with bremsstrahlung
neglected in both cases. The solid curves give the numerical results with
bremsstrahlung taken into account.
}\label{f:bremtrn}
\end{figure}

We have calculated the turnover frequency as a function of $\Theta$ and $B$ for
the remaining parameters fixed, see Figures \ref{f:bremtrn}a, b, respectively.
As expected, $\nu_{\rm t}$ is determined mostly by bremsstrahlung at low
temperatures and weak magnetic fields. However, though bremsstrahlung can have
a negligible effect on the value of $\nu_{\rm t}$, the total emitted power can
still be dominated by it, see Section \ref{applications} below, and, e.g.\
Figure 3 in Z86. On the other hand, if bremsstrahlung self-absorption dominates
over the synchrotron one, bremsstrahlung emission will also dominate over the
CS emission, except in the case of spectra of the latter emission being harder
than the bremsstrahlung spectrum.

\section{Comptonization of synchrotron photons}
\label{s:compt}

Synchrotron photons produced in a plasma will, in general, be Compton scattered
by the plasma electrons. In the case of a thermal plasma, the synchrotron
spectrum is usually self-absorbed up to a high value of $\nu/\nu_{\rm c}$
(Section \ref{ss:turn}). In that case, the self-absorbed synchrotron
spectrum is rather narrow, consisting of a hard Rayleigh-Jeans spectrum at
$\nu\la \nu_{\rm t}$ (in which regime scattering can be neglected, $\tau \gg
\ts$, $\tau \gg 1$), and a fast-declining tail of the optically-thin
synchrotron spectrum at $\nu\ga \nu_{\rm t}\gg \nu_{\rm c}$ (where absorption
can be neglected,  $\tau \ll \ts$, $\tau \ll 1$). Photons in that spectrum then
serve as seed photons for Comptonization. For that process, it is usually
sufficient to treat the self-absorbed synchrotron photons as monochromatic at
$\nu_{\rm t}$ (see, e.g., Monte Carlo simulations in Z86). Also, for magnetic
fields, $B \la 10^9$ G, and electron temperatures, $kT \ga 10$ keV,  expected
in compact sources, $h\nu_{\rm t}\ll kT$, and photons at $\nu\sim \nu_{\rm t}$
gain energy in the scattering process.

To treat thermal Comptonization, we use the method of Zdziarski (1985,
hereafter Z85), as applied to thermal synchrotron sources in Z86. This method
gives an approximate form of Comptonized spectra, reproducing relatively
accurately the energy spectral index, $\alpha$, and the total power in the
scattered spectrum (see comparison with Monte Carlo results in Z86). On the
other hand, it gives a relatively inaccurate shape of the high-energy cutoff of
the scattered spectrum (see, e.g., Poutanen \& Svensson 1996), which, however,
is of negligible importance for our applications.

We consider a homogeneous and isotropic source characterized by $\ts$, $\Theta$
and $B$. The flux in scattered photons can be approximated as a sum of a
cut-off power law and a Wien component,
\begin{equation}
\label{density}
F_{\rm C} (x) = N_{\rm P} \left( \frac{x}{\Theta}\right)^{-\alpha}
{\rm e}^{-x/\Theta} +  N_{\rm W} \left(\frac{x}{\Theta}\right)^3 {\rm
e}^{-x/\Theta},
\end{equation}
where $x\equiv h\nu/ m_{\rm e} c^2$ is a dimensionless photon energy,
and $N_{\rm P}$, $N_{\rm W}$ are normalizations of the power-law and Wien
components, respectively.

For $\ts \la 2$, $\alpha$ can be approximately given by,
\begin{equation}
\label{indeks} \alpha = - \frac{\ln P_{\rm sc}}{\ln A}, \qquad A \equiv 1 +
\left\langle \frac{\Delta \nu}{\nu}\right\rangle \approx 1 + 4 \Theta + 16
\Theta^2, \end{equation}
where $A$ is the average photon energy amplification per scattering and $P_{\rm
sc}$ is the scattering probability averaged over the source volume.  Note that
$\alpha>0$ in general.  In spherical geometry (Osterbrock 1974),
\begin{eqnarray} \label{sferpraw} \lefteqn{ P_{\rm sc} = 1 -
\frac{3}{8 \ts^3} \left[ 2 \ts^2 - 1 + {\rm e}^{-2 \ts} (2 \ts +1) \right]}
\nonumber \\ \lefteqn{ \qquad \rightarrow \cases{ {3\ts\over 4}, &$\ts\ll
1$;\cr 1-{3\over 4\ts}, &$\ts\gg 1$,\cr} } \end{eqnarray}
where $\ts$ is the Thomson optical depth along the radius.  We have found a
power-law approximation to $\alpha(\ts,\Theta)$ valid at $0.5\la \ts\la 2$ and
$\Theta\la 0.4$,
\begin{equation} \label{alpha} \alpha\approx {3\over 19} {1\over
\ts^{4/5} \Theta}.  \end{equation}
See Zdziarski et al.\ (1994) for $P_{\rm sc}$ in the slab geometry. In general,
different prescriptions for $\alpha(\tau,\Theta)$ can be used, e.g., in order
to account for a different geometry or to increase the accuracy, without
affecting the validity of the results presented below in this section. On the
other hand, some equations in Section \ref{applications} utilize equation
(\ref{alpha}), and using a different $\alpha(\tau,\Theta)$ would somewhat
change those formulae.

Note that at $\ts\ll 1$, scattering profiles corresponding to subsequent
scatterings are visible in the Compton spectrum, and a power law is no longer a
good approximation to the shape of the spectrum below the cutoff (e.g., Fig.\
2b in Z86). Still, the power-law description presented here gives the shape
averaged over the scattering orders, and an integral over that approximate form
gives a fair approximation to the total luminosity (see, e.g., Monte Carlo
simulations in Z86). On the other hand, at $\ts\ga 2$--3, Comptonization can be
described by means of a kinetic, Fokker-Planck, equation (Sunyaev \& Titarchuk
1980; see e.g.\ Lightman \& Zdziarski 1987 for relativistic and low-$\ts$
corrections). The two solutions can be matched at $\ts\sim 2$ (Z85). The
remaining results presented in this Section can still be used at $\ts\ga 2$--3,
except for a different prescription for $\alpha$. In particular, the ratio of
$N_{\rm W}/N_{\rm P}$ given by Z85,
\begin{equation}
\label{relats}
\frac{N_{\rm W}}{N_{\rm P}} = \frac{\Gamma(\alpha)}{\Gamma(2 \alpha +3)}
P_{\rm sc},
\end{equation}
where $\Gamma$ is Euler's gamma function, and which uses a result of Sunyev \&
Titarchuk (1980), is a good approximation in both the optically-thin and
optically-thick regimes (see, e.g., comparison with Monte Carlo results in
Z86).

We need then to normalize the flux in the Compton-scattered spectrum, $F_{\rm
C}(x)$, with respect to that in the Rayleigh-Jeans spectrum, $F_{\rm
RJ}(x)$. In general, the self-absorbed synchrotron spectrum at its peak around
$x_{\rm t}$ ($\equiv h\nu_{\rm t}/m_{\rm e} c^2$) is somewhat above an
extrapolation of the power-law component of the Compton spectrum, $F_{\rm
C}(x)$ (which follows from photon conservation, Z85). Z86 finds the following
phenomenological relation providing a good approximation to the relative
normalization,
\begin{equation}
\label{normalizacja}
F_{\rm C}(x_{\rm t}) = \varphi F_{\rm RJ} (x_{\rm t}), \quad
\varphi(\Theta)\approx
\frac{1+ (2\Theta)^2}{1+ 10 (2 \Theta)^2}.
\end{equation}
This is valid for $x_{\rm t}\ll \Theta$, which condition we assume
hereafter. The flux of the Rayleigh-Jeans spectrum is given by,
\begin{equation}
\label{rjlum}
F_{\rm RJ} (x) = \upi I_{\rm RJ} (x),  \quad
I_{RJ} (x) = \frac{2 m_{\rm e} c^3 \Theta}{\lambda_{\rm C}^3} x^2,
\end{equation}
where $I_{\rm RJ}$ is the specific intensity and $\lambda_{\rm C}\simeq
2.426\times
10^{-10}$ cm is the electron Compton wavelength.  Then,
\begin{equation}
\label{np}
N_{\rm P} = \upi \varphi I_{RJ}(x_{\rm t}) (x_{\rm t} /\Theta)^\alpha .
\end{equation}
The luminosity due to the CS emission is then given by,
\begin{equation}
\label{intens} \label{calamoc}
L_{\rm CS}= {\cal A} \left[ \int_0^{x_{\rm t}} {\rm d}x \; F_{\rm RJ}(x)
+  \int^\infty_{x_{\rm t}} {\rm d} x \; F_{\rm C} (x) \right].
\end{equation}
where ${\cal A}$ is the source area, and which can be integrated to,
\begin{eqnarray}
\label{specint}
\lefteqn{ L_{\rm CS}= {\cal A} {2\upi m_{\rm e} c^3 \Theta x_{\rm t}^3 \over 3
\lambda_{\rm C}^3}
\left\{ 1 + 3\varphi \left( x_{\rm t}\over \Theta\right)^{\alpha -1} \times
\right. }   \nonumber\\
\lefteqn{\left. \qquad \left[ \Gamma\left( 1-\alpha,
{ x_{\rm t}\over \Theta}\right) +
\frac{6 \Gamma(\alpha) P_{\rm
sc} }{\Gamma(2 \alpha +3)} \right] \right\},}
\end{eqnarray}
where the incomplete gamma function is well approximated for $x_{\rm t}\ll
\Theta$ by,
\begin{eqnarray}
\label{Gamma}
\lefteqn{ \Gamma\left( 1-\alpha,
{ x_{\rm t}\over \Theta}\right)\simeq } \nonumber\\
\lefteqn{  \cases{
 \ln{\Theta\over x_{\rm t}} -\gamma_{\rm E}, & $|\alpha-1|<10^{-3}$;\cr
{\Theta\over x_{\rm t}}-\ln{\Theta\over x_{\rm t}} -1+\gamma_{\rm E}, &
$|\alpha-2|<10^{-3}$;\cr
{1\over \alpha-1} \left(\Theta\over x_{\rm t}\right)^{\alpha-1}, & $\alpha>
2.99$;\cr
\Gamma(1-\alpha) -{(x_{\rm t}/\Theta)^{1-\alpha}\over 1-\alpha} +{(x_{\rm
t}/\Theta)^{2-\alpha}\over 2-\alpha} , & otherwise. }}
\end{eqnarray}
At $x_{\rm t}/\Theta=0.01$, the maximum relative error of this approximation
of $\sim 0.02$ occurs around $\alpha\simeq 3$. The relative error declines
rapidly at lower values of $x_{\rm t}/\Theta$ and $\alpha$; e.g., it is
$<0.001$ at $x_{\rm t}/\Theta=0.01$ and $\alpha\leq 2.9$, and $<0.002$ at
$x_{\rm t}/\Theta=10^{-3}$ at any $\alpha$.

In an astrophysically important case of $\alpha\sim 0.4$--0.9 (e.g.\
Gierli\'nski et al.\ 1997; Zdziarski, Lubi\'nski \& Smith 1999), the main
contribution to the luminosity comes from the high-energy end of the spectrum,
and the Wien component is relatively unimportant. Then, we obtain an
approximation valid within a factor of $\la 2$ at $x_{\rm t}/\Theta\la
10^{-5}$,
\begin{equation}
\label{hardL}
L_{\rm CS} \approx {8\upi^2 m_{\rm e} c^3 R^2 \over (1-\alpha)
\lambda_{\rm C}^3}  x_{\rm t}^{2+\alpha} \varphi \Theta^{2-\alpha},
\end{equation}
where we assumed a spherical geometry. On the other hand, for soft spectra,
with $\alpha\ga 1.1$, we get an approximation valid within $\sim 30$ per cent,
\begin{equation}
\label{softL}
L_{\rm CS} \approx {8\upi^2 m_{\rm e} c^3 R^2 \over 3 \lambda_{\rm C}^3}
\left(1+{3\varphi\over \alpha-1} \right)  x_{\rm t}^3 \Theta .
\end{equation}
For order-of-magnitude estimates, we can then substitute $x_{\rm t}$ of
equation (\ref{nu_ratio}) in equations (\ref{hardL})-(\ref{softL}). This yields
$L_{\rm CS}\propto R^2 \Theta^{3.78-0.11\alpha} B^{1.82+0.91\alpha}$ and
$\propto R^2 \Theta^{3.66} B^{2.73}$ for $0.4\la \alpha\la 0.9$ and $\alpha\ga
1.1$, respectively.

We stress that the above formulae for the Comptonization spectral shape and the
corresponding expressions for the luminosity are mutually consistent. This is
much preferable to computing Comptonization luminosity {\it independently\/} of
its corresponding spectrum by multiplying the power in a self-absorbed
synchrotron spectrum by an amplification factor of Comptonization (e.g.\ of
Dermer, Liang \& Canfield 1991), as often done in studies of advective flows.

\section{Applications to accreting black holes}
\label{applications}

\begin{figure}\begin{center}\leavevmode
\epsfxsize=7.2cm \epsfbox{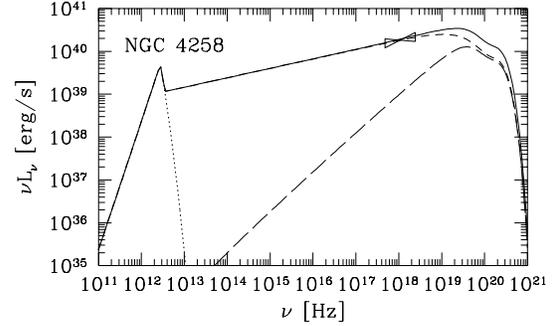}\end{center}
\caption{
An example of the spectrum of a magnetized plasma, for $R= 9\times 10^{13}$ cm,
$B= 2.2 \times 10^3$ G, $\tau_{\rm T}= 0.47$ and $\Theta= 0.4$. The dotted,
short-dashed and
long-dashed curves represent spectra due to synchrotron emission, its
Comptonization and Comptonized bremsstrahlung, respectively. The solid curve
represents the sum of those components. This spectrum also represents a model
of a hot accretion flow fitted to the X-ray data for NGC 4258 (bow-tie,
Makishima et al.\ 1994), see Section \ref{hot_ions}. }
\label{f:ngc}
\end{figure}

Figure \ref{f:ngc} shows an example of the CS spectrum for parameters typical
to low-luminosity AGNs.  The spectrum has the Rayleigh-Jeans shape below the
turnover frequency, it is a power law due to Comptonization above that
frequency, and then it has a thermal high-energy cutoff.  We also show the
contribution due to Comptonized bremsstrahlung, which is moderately important
for the chosen parameters.  We have computed the latter spectrum by treating
the spectrum of equation \odn{density} as Green's function for Comptonization
and then integrating over the seed spectrum of optically-thin bremsstrahlung.
The spectrum is computed assuming spherical geometry.  More examples of spectra
from magnetized plasmas are given, e.g.\ in Z86.

For given $\Theta$, the Rayleigh-Jeans part of the spectrum is independent of
$B$, and its spectral luminosity is $L_{\rm RJ}(\nu) \propto R^2$. On the other
hand, the CS spectrum obeys $L_{\rm CS} (\nu) \propto R^2 \nu_{\rm
t}^{2+\alpha}$ [equation (\ref{np})], i.e., its normalization increases quickly
with increasing $B$ via the dependence of the turnover frequency on $B$. When
self-absorption is dominated by the synchrotron process, $\nu_{\rm t}$ is
roughly $\propto B$ with no explicit dependence on $R$, see equation
(\ref{nu_ratio}). The total luminosity, $L_{\rm CS}$, follows the same
dependence if $\alpha\la 1$, see equation (\ref{hardL}), and $L_{\rm CS}
\propto R^2 \nu_{\rm t}^3$ for $\alpha\ga 1$, see equation (\ref{softL}). The
shape of the Comptonized bremsstrahlung spectrum is almost independent of
$B$ (except for a very weak dependence of Comptonization on $\nu_{\rm t}$).

The bremsstrahlung luminosity can be constrained from below as
\begin{equation}
\label{lcb}
\frac{L_{\rm CB}}{L_{\rm E}} > 1.2 \times 10^{-6} \Theta^{1/2} \ts^2 r\approx
1.2 \times 10^{-8} \alpha^{-5/2} \Theta^{-2} r,
\end{equation}
where $r\equiv R/R_{\rm g}$, $R_{\rm g}\equiv GM/c^2$ is the gravitational
radius, $L_{\rm E}\equiv 4\upi \mu_{\rm e} GM m_{\rm p} c/\sigma_{\rm T}\approx
1.5\times 10^{38} m$ erg s$^{-1}$ is the Eddington luminosity, $\mu_{\rm
e}=2/(1+X)$ is the mean electron molecular weight, and $X$ ($\approx 0.7$) is
the H mass fraction, and $m\equiv M/{\rm M}_\odot$.  The right-hand-side
expressions neglect Comptonization and relativistic corrections, which effects
will increase the actual $L_{\rm CB}$.  The last expression uses the
approximation of equation (\ref{alpha}) to $\alpha(\ts,\Theta)$.  Note that
$L_{\rm CB}$ represents a strict lower limit to the luminosity of a source with
given $\ts$, $\Theta$ and size.  Thus, the luminosity of weak sources is
dominated by $L_{\rm CB}$, with $L_{\rm CS}$ being negligible.  Note, however,
that usually the plasma parameters depend on luminosity, which dependence
should be taken into account when determining the luminosity below which
bremsstrahlung dominates.

\begin{figure}\begin{center}\noindent
\epsfxsize=7.2cm \epsfbox{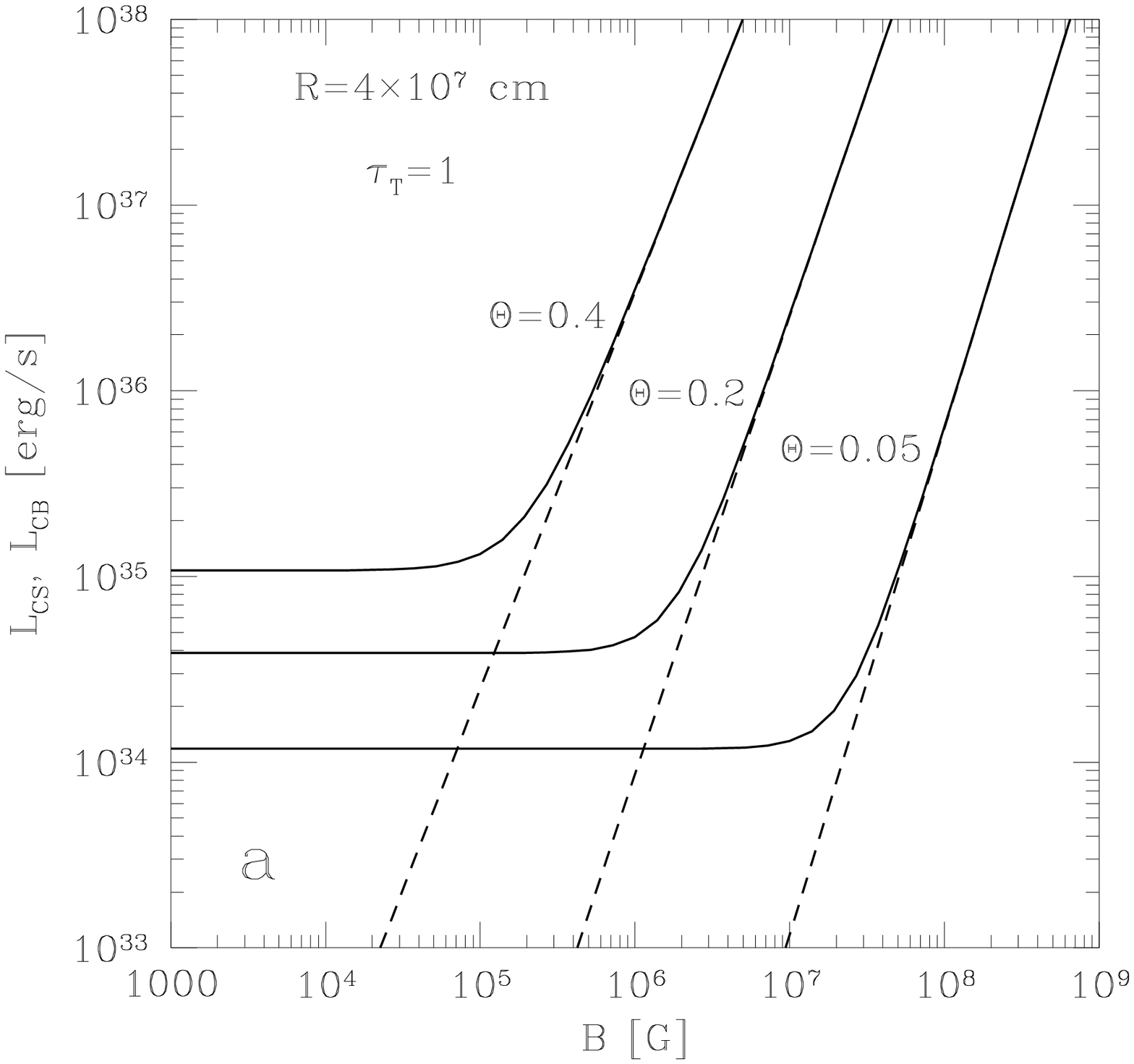}\end{center}\nobreak
\begin{center}\noindent
\epsfxsize=7.2cm \epsfbox{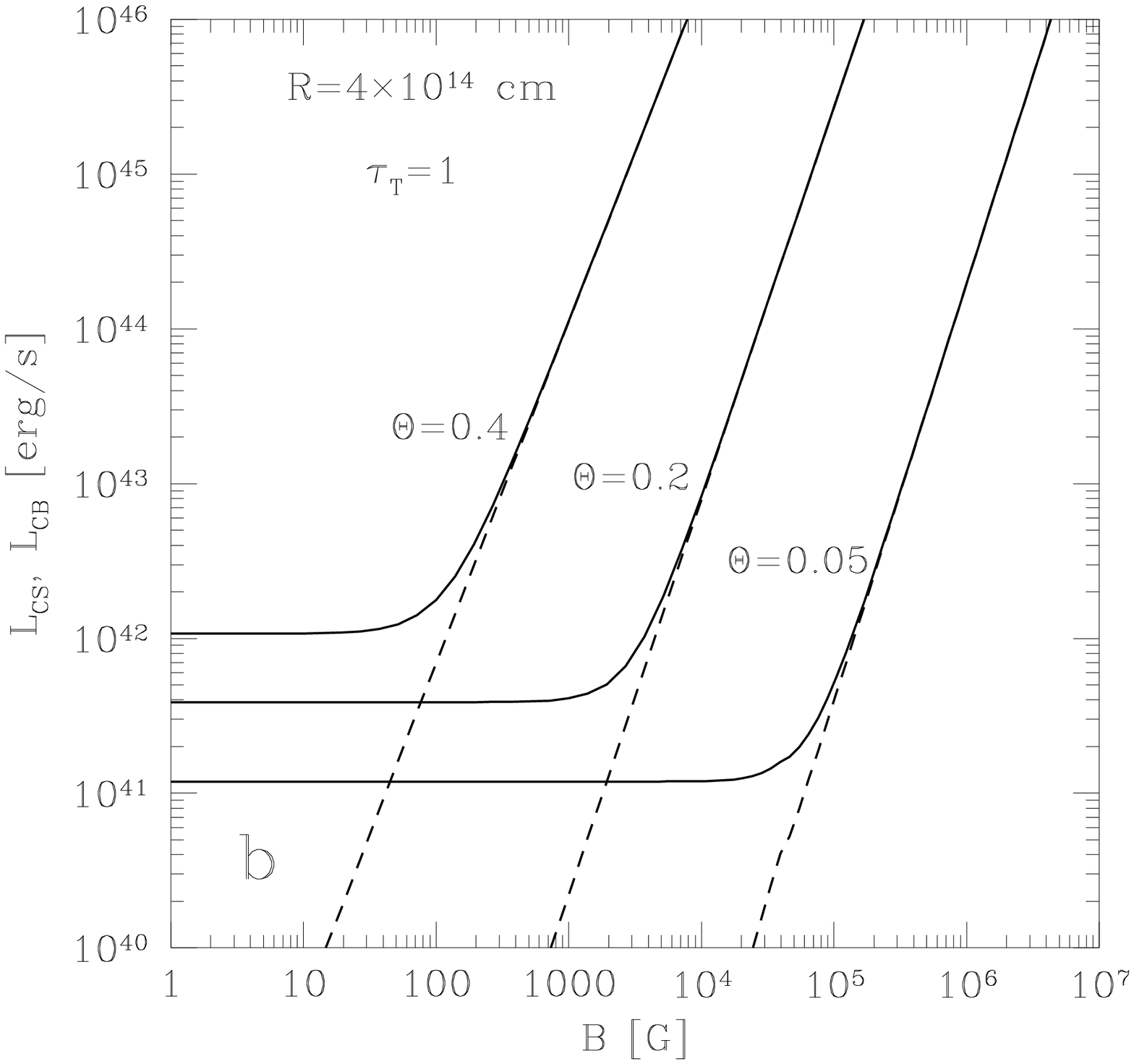}\end{center}
\caption{
Example dependences of the plasma luminosity on $B$ for (a) $R=4\times
10^7$ cm and (b) $R=4\times 10^{14}$ cm for three values of temperature. The
flat parts correspond to dominant bremsstrahlung, and dashed curves give the
luminosity in the CS component alone.
}\label{f:power}
\end{figure}

Figure \ref{f:power} shows some example dependences of the total $L$ on $B$ at
$\Theta=0.05$--0.4 for $R=4\times 10^7$ cm and $4\times 10^{14}$ cm
(corresponding to $\sim 30R_{\rm g}$ for 10\msol\ and $10^8$\msol\ black-hole
mass, respectively).  The flat parts at low values of $B$ correspond to the
dominant bremsstrahlung, with its relative importance increasing with
decreasing $\Theta$.

Hereafter, we use the symbols $L$ and $\eta$ for the total luminosity of a hot
plasma in a source, including {\it all\/} radiative processes, and for the
corresponding Eddington ratio, $L/L_{\rm E}$, respectively.  This $L$ then
corresponds to the one observed (excluding components not originating in the
hot plasma, e.g., a disc blackbody).  We then compare observed values of $L$
with the CS luminosity assuming some specific prescriptions for the magnetic
field strength in an accretion flow.  Typically, the maximum possible magnetic
field strength corresponds to equipartition between the pressure or energy
density of the field and of the gas and radiation in the source (Galeev, Rosner
\& Vaiana 1979).

\subsection{Two-temperature accretion flows}
\label{hot_ions}

Here, we consider the CS emission from optically thin, two-temperature
accretion flow. In inner parts of hot accretion flows, the ion temperature,
$T_{\rm i}$, is typically much higher than the electron temperature (e.g.\
Shapiro, Lightman \& Eardley 1976; NY95), and the ion energy
density is much higher than that of both electrons and radiation. Then, energy
density equipartition corresponds to
\begin{equation} \label{field}
\frac{B^2}{8 \upi} =
{3 \over 2}n_{\rm e}{\mu_{\rm e}\over \mu_{\rm i}} kT_{\rm i},
\end{equation}
where $\mu_{\rm i}= 4/(1+3X)$ is the mean ion molecular weight. We note that
pressure equipartition would result is a slightly different condition, and that
the magnetic-field pressure, $B^2/8\upi$,  is sometimes assumed to equal
$B^2/24 \upi$ (e.g., Mahadevan 1997).

Maximum possible ion temperatures are sub-virial, and detailed accretion flow
models show $kT_{\rm i}\approx \delta m_{\rm p} c^2/r$, where $\delta\ll 1$
(Shapiro et al.\ 1976; NY95).  In particular, $\delta$ constant with $r$ and
dependent on the flow parameters is obtained in the self-similar
advection-dominated solution of NY95.  Close to the maximum
possible accretion rate of the hot flow, $\delta\sim 0.2$ is a typical value
[see equation (2.16) in NY95], which we adopt in numerical calculations below.
However, the self-similar solution breaks down in the inner flow, which results
in values of $T_{\rm i}$ much lower than that given by the self-similar
solution, e.g., by a factor of $\sim 10$ at $r=6$ (Chen, Abramowicz \& Lasota
1997).  Thus, we can obtain an upper limit on the equipartition magnetic field
strength by assuming that it follows the self-similar solution above some
radius, $r_0\sim 20$, while it remains constant at $r<r_0$,
\begin{equation}\label{B} B\simeq 7.3\times 10^8 \,{\rm G} \left(\delta
\ts\over m \right)^{1/2} \times\cases{ 1/r_0, & $r\leq r_0$;\cr 1/r, &
$r>r_0$.} \end{equation} This typically gives $B\la 10^7$ G and $\la 10^4$ G in
inner regions of binary sources and AGNs, respectively.

Then, the CS luminosity can be constrained from above by a sum of the
contribution from the inner region (assumed here to be spherical) with radius
$r_0$, and a contribution from the outer region $r>r_0$, obtained here by
radial integration all $r_0$ to infinity (for which we assumed a planar
geometry). We use here the upper limit on $B$ of equation (\ref{B}) and assume
$\alpha$ and $\Theta$ constant through the flow. The latter assumption
maximizes $L_{\rm CS}$ of given $\alpha$ and $\Theta$ since both $\ts$ and
$\Theta$ will, in fact, decrease with $r$ in an accretion flow. (We also note
that the maximum of dissipation per $\ln r$ in the Schwarzschild metric occurs
at $r\sim r_0\sim 20$.) For $0.4\la\alpha\la 0.9$, the resulting upper limit is
\begin{eqnarray}\label{hard_ratio}
\lefteqn{{L_{\rm CS}\over L_{\rm E}}\approx
{43 \times 10^{-2.83 \alpha} \delta^{0.91+0.46 \alpha}
\Theta^{2.64-0.68\alpha} \varphi\over
(m r_0^2)^{0.46\alpha-0.09} (1-\alpha)\alpha^{0.63\alpha+1.26}}
\times}\nonumber\\
\lefteqn{\qquad \quad \cases{1 & for inner region;\cr
(0.91\alpha-0.18)^{-1} & for outer region,} }
\end{eqnarray}
where we used equations (\ref{alpha}), (\ref{hardL}), (\ref{B}) and
$\varphi\sim 0.5$.
The total $L_{\rm CS}$ corresponds to the sum of the contributions from the two
regions. In the case of $\alpha\ga 1.1$, we have an upper limit on the
luminosity of
\begin{eqnarray}\label{soft_ratio}
{L_{\rm CS}\over L_{\rm E}}\approx \frac{ 0.021
\delta^{1.37}\Theta^{1.96}}{m^{0.37}r_0^{0.73}\alpha^{1.89}}
\left(1+\frac{3 \varphi}{\alpha-1}\right)
\times \nonumber\\
\cases{1 & for inner region;\cr 1.37 & for outer region,}
\end{eqnarray}
which is based on equation (\ref{softL}).

We see from equations (\ref{hard_ratio})-(\ref{soft_ratio}) that the predicted
Eddington ratio decreases with the mass. Then, if the CS process dominates, the
predicted X-ray spectra would harden with increasing $M$ at a given $L_{\rm
CS}/L_{\rm E}$. In fact, the {\it opposite\/} trend is observed; black-hole
binaries have X-ray spectra harder on average than those of AGNs (e.g.\
Zdziarski et al.\ 1999), implying that either the CS process does not dominate
the X-ray spectra of those objects or (less likely) the Eddington ratio is much
lower on average in AGNs than in black-hole binaries.

We note that this conclusion differs from a statement in NY95 that hot
accretion flows with CS cooling are effectively mass-invariant.  In particular,
we have been unable to explain their Fig.\ 4, in which flows onto black holes
have radial dependences of both the temperature at the critical (i.e., maximum
possible) accretion rate and that rate itself (in Eddington units) virtually
identical for $m=10$ and $m=10^8$.  This would then imply very similar X-ray
spectra in both cases. On the other hand, we have found a good agreement of
our results with the dependences on $m$ in the results of Mahadevan
(1997).

\begin{figure}\begin{center}\noindent
\epsfxsize=7.2cm \epsfbox{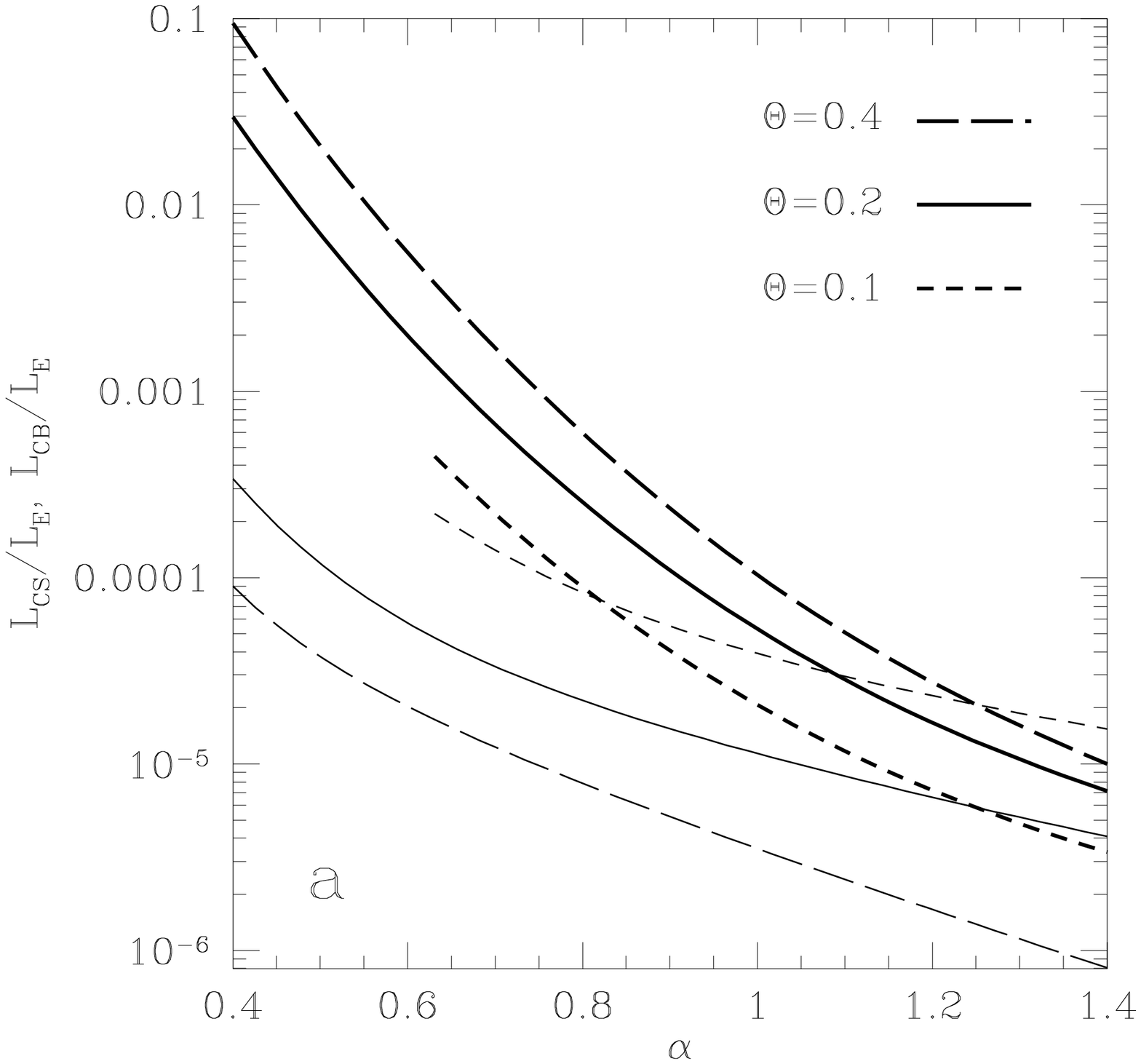}\end{center}\nobreak
\begin{center}\noindent
\epsfxsize=7.2cm \epsfbox{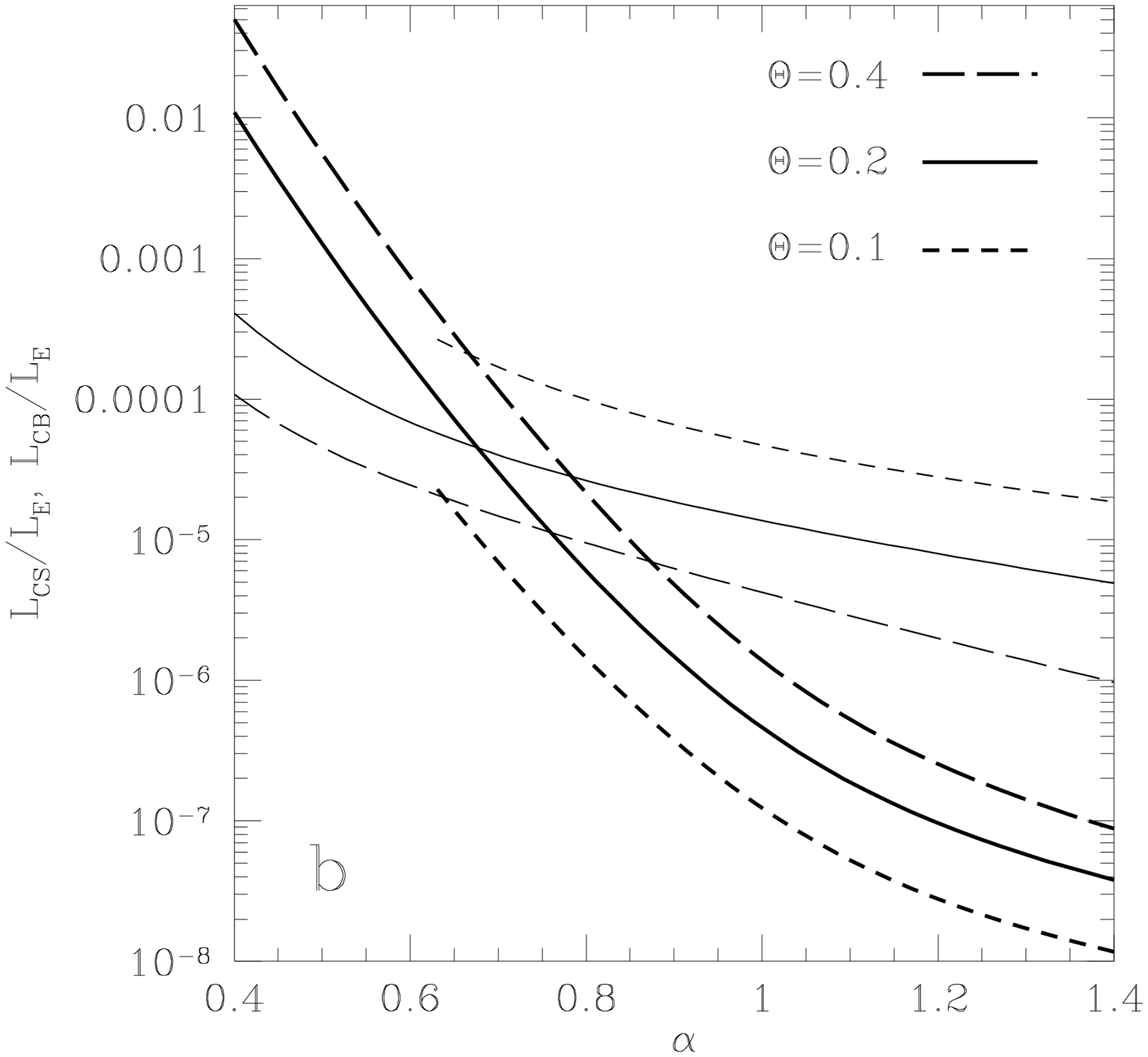}\end{center}
\caption{The Eddington ratio as a function of the spectral index of CS
radiation and for 3 values of electron temperature for masses (a)
$m=10$, and (b) $m=10^8$. The heavy and thin curves show the luminosity in the
CS radiation and in Comptonized bremsstrahlung,
respectively. See text for other assumptions. The curves are shown only for
values of $\alpha$ corresponding to $\ts\leq 3$, as for larger $\ts$ our
formulae for $\alpha$ (Section \ref{s:compt}) break down.
}\label{f:ratio}
\end{figure}

We find that, under conditions typical to compact sources, the Comptonized
bremsstrahlung luminosity, $L_{\rm CB}$, is often comparable to or higher than
$L_{\rm CS}$.  Figure \ref{f:ratio} shows the results of numerical calculations
of the Eddington ratio for both CS and bremsstrahlung radiation as a function
of the spectral index of the CS radiation, $\alpha$, for $m=10$ and $10^8$.  To
enable direct comparison of those two components, we show the luminosity from
the inner region (with constant $B$) only, assuming $r_0=20$ and $\delta=0.2$.
Thick and thin curves give $L_{\rm CS}$ and $L_{\rm CB}$ [see equation
(\ref{lcb}) with $r=r_0$], respectively.  The relative importance of
bremsstrahlung increases with increasing mass and size, e.g.\ $L_{\rm
CB}/L_{\rm CS}\propto m^{0.46\alpha-0.09} r_0^{0.91\alpha +0.82}$ and $\propto
m^{0.37} r_0^{1.73}$ for $\alpha\la 0.9$ and $\ga 1.1$, respectively.  Note
that $L_{\rm CB}$ decreases with increasing $\Theta$ at a constant $\alpha$,
which is an effect of a strong dependence of $L_{\rm CB}$ on $\ts$ (increasing
with decreasing $\Theta$) and a weak dependence of $L_{\rm CB}$ on temperature.
Also, an important conclusion from Figure \ref{f:ratio} is that Eddington
ratios $\ga 0.01$ can be obtained from this process only for very hard spectra
and high electron temperatures, with this constraint being significantly
stronger in the case of AGNs.

\begin{figure}
\begin{center} \noindent
\epsfxsize=7.2cm \epsfbox{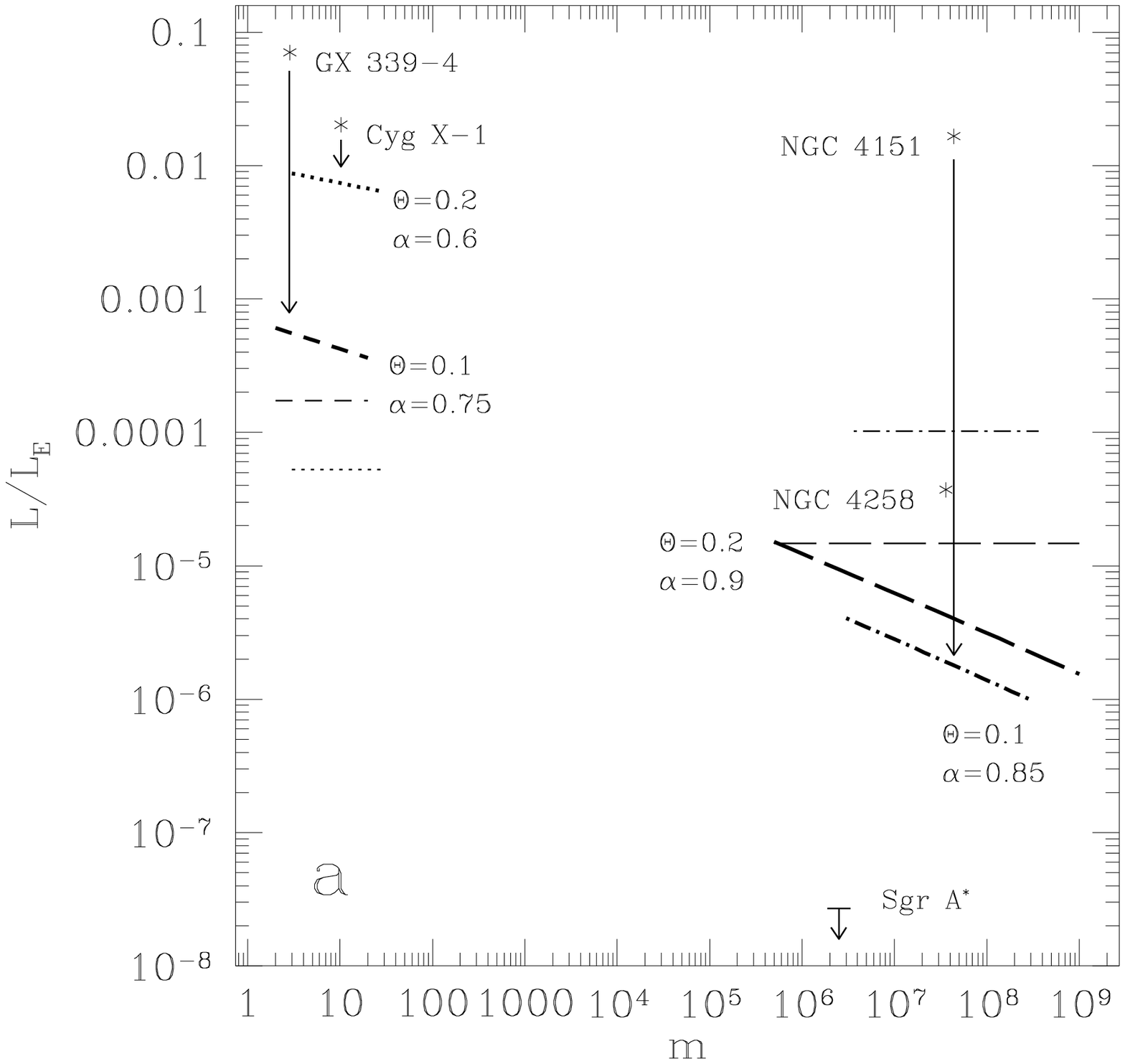}\end{center}\nobreak
\begin{center}\noindent
\epsfxsize=7.2cm \epsfbox{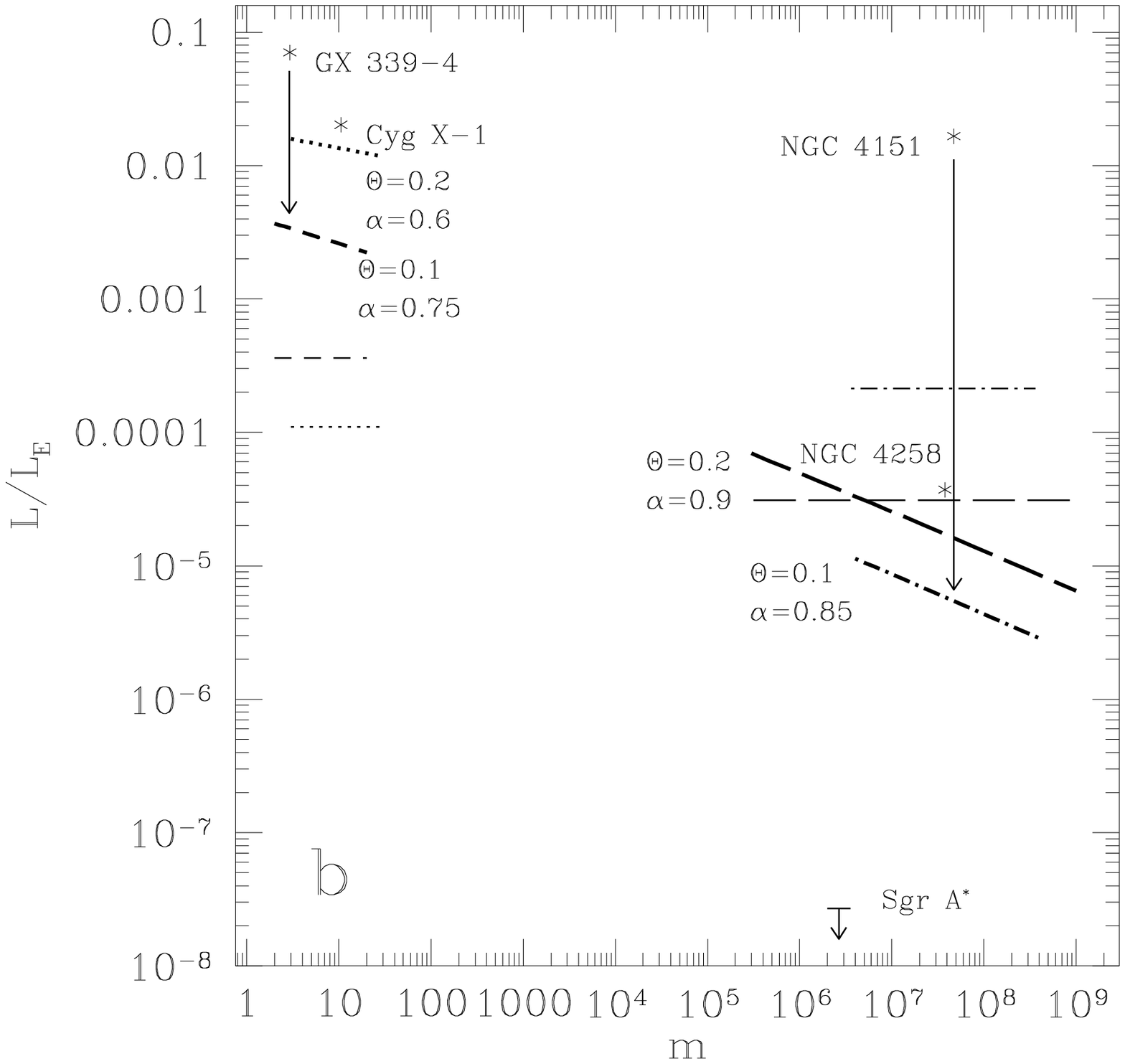}\end{center}
\caption{Comparison of Eddington ratios inferred from observations (asterisks,
except for Sgr A$^*$, where an upper limit is shown) with the model upper
limits from the CS emission (heavy curves labeled by values of $\Theta$ and
$\alpha$) for a range of black-hole masses.
Vertical arrows connect observed values with the corresponding CS model upper
limits, except for NGC 4258 and Sgr A$^*$, for which no sufficient spectral
data exist.  On the other hand, the model curves for $\Theta=0.2$,
$\alpha=0.9$, corresponding to the average spectrum of Seyfert 1s, have no
corresponding observational point. Thin horizontal lines show the Comptonized
bremsstrahlung emission. The magnetic field energy density is in equipartition
with (a) ions in a hot accretion flow, and (b) with radiation energy density
times $10\,c/v_{\rm A}$ in a patchy corona, see Sections \ref{hot_ions} and
\ref{dissipation}, respectively.  }
\label{comparison}
\end{figure}

Figure \ref{comparison}a compares the upper limits on $L_{\rm CS}/L_{\rm E}$
with values of $L/L_{\rm E}$ inferred from observations for a number of objects
and with the corresponding luminosity from bremsstrahlung. The shown values of
$L_{\rm CS}$ include the contributions from both the inner and the outer region
of the flow [equation (\ref{B})] for $r_0=20$ and $\delta=0.2$. On the other
hand, the shown values of $L_{\rm CB}$ are for emission from within $r_0$ only,
and the expected actual $L_{\rm CB}$ will be by a factor of $\sim 2$ larger.
For Cyg X-1 in the hard state, we used the brightest spectrum of Gierli\'nski
et al.\ (1997), with $\alpha=0.6$, $\Theta=0.2$, and $L\approx 3\times 10^{37}$
erg s$^{-1}$ (assuming the distance of $D=2$ kpc, Massey, Johnson \&
DeGioia-Eastwood 1995; Malysheva 1997) and  $m=10$. For GX 339--4 in the hard
state, we used $\alpha=0.75$, $\Theta=0.1$, $L\approx 3\times 10^{37}$ erg
s$^{-1}$, $D=4$ kpc and $m=3$ (Zdziarski et al.\ 1998).  For NGC 4151, we used
the brightest spectra observed by {\it Ginga\/} and {\it CGRO}/OSSE (Zdziarski,
Johnson \& Magdziarz 1996), for which $\alpha=0.85$, $\Theta=0.1$, and
$L\approx 9\times 10^{43}$ erg s$^{-1}$,  assuming $D=16.5$ Mpc (corresponding
to $H_0=75$ km s$^{-1}$ Mpc$^{-1}$), and $m=4\times 10^{7}$ (Clavel et al.\
1987). For NGC 4258 we adopted $L=2\times 10^{41}$ erg s$^{-1}$ from an
extrapolation of the 2--10 keV luminosity of $L_{\rm 2-10keV}=3.1\times
10^{40}$ erg s$^{-1}$ with $\alpha=0.78$ (Makishima et al.\ 1994) up to 200
keV, $D=6.4$ Mpc and $m=3.6\times 10^7$ (Miyoshi et al.\ 1995). Finally, for
Sagitarius A$^*$ we assume an upper limit of $L<10^{37}$ erg s$^{-1}$ (Narayan
et al.\ 1998), and $m=2.5\times 10^6$ (Eckart \& Genzel 1996). Finally, we show
$L_{\rm CS}$ for the average parameters of Seyfert-1 spectra, $\alpha=0.9$,
$\Theta=0.2$ (e.g.\ Zdziarski 1999).

We see that only in the case of Cyg X-1, with its relatively hard spectrum, the
CS emission can contribute substantially, at $\la 30$ per cent, to the total
luminosity.  Given uncertainties of our model, e.g., in the value of $\delta$,
it is in principle possible that the CS process can account for most of the
emission of Cyg X-1.  On the other hand, we have adopted here assumptions
maximizing synchrotron emission and we consider it more likely that the CS
process is negligible in Cyg X-1.  This appears to be supported by the
similarity between X-ray spectra and the patterns of time variability between
Cyg X-1 and the other black-hole binary considered by us, GX 339--4 (Miyamoto
et al.\ 1992; Zdziarski et al.\ 1998).  In the latter object, $L_{\rm CS}$
clearly provides a negligible contribution to $L$, as shown by Zdziarski et
al.\ (1998).  Furthermore, a remarkable similarity exists between X-ray spectra
of black-hole binaries and Seyfert 1s (e.g.\ Zdziarski et al.\ 1999).  For the
latter, $L_{\rm CS}/L_{\rm E}\sim 10^{-5}$--$10^{-6}$ are found, whereas the
typical Eddington ratios of those objects are likely to be $\sim 0.01$ (e.g.,
Peterson 1997), i.e., 3--4 orders of magnitude more.  Concluding, CS emission
in the assumed hot-disc geometry is unlikely to be responsible for the observed
X-ray spectra of luminous black-hole sources.  To explain those spectra, an
additional source of soft seed photons is required.  Such seed photons are
naturally provided by blackbody emission of some cold medium, e.g\ an
optically-thick accretion disc or cold blobs, co-existing with the hot flow.

On the other hand, the CS process can clearly be important in weaker sources,
e.g., NGC 4258.  We have compared predictions of our model with the 2--10 keV
spectral index and luminosity (see above) of this object.  We have obtained a
good fit to the data, assuming $\delta=0.2$, $\Theta=0.4$ and taking into
account emission within $r_{0}=17$, as shown in Figure \ref{f:ngc}. This yields
$L_{\rm CS}=1.8\times 10^{41}$ erg ${\rm s^{-1}}$ and $L_{\rm CB}=4\times
10^{40}$ erg ${\rm s^{-1}}$.  A similar spectrum from an advection disc model
was obtained by Lasota et al.\ (1996).

For $\Theta \approx 0.2$ (typical for luminous sources), we could reproduce the
$L_{\rm 2-10keV}$ of NGC 4258, but the 2--10 keV spectrum was dominated by
bremsstrahlung, i.e., much harder than the observed one. The relatively high
temperature, $\Theta\simeq 0.4$, required in our model, is, in fact, consistent
with predictions of hot accretion flow models (e.g. NY95).

As stated above, Comptonized bremsstrahlung, with $L_{\rm CB}/L_{\rm E}$
independent of $M$, provides the minimum possible luminosity of a plasma with
given $\ts$, $\Theta$ and $r$.  This is shown by horizontal lines in Figure
\ref{comparison}.  Thus, sources with lower luminosities cannot have plasma
parameters characteristic of luminous sources, which is indeed consistent with
hot accretion flow models, in which $\ts$ quickly decreases with decreasing $L$
(e.g.\ NY95).  On the other hand, CS emission can still be important in a
certain energy range above the turnover energy even if $L_{\rm CB}>L_{\rm CS}$.
This may be the case, e.g., in Sgr A$^*$ (Narayan et al.\ 1998).

\subsection{Active coronae}
\label{coronae}

\subsubsection{Equipartition with radiation energy density}
\label{radiation}

Di Matteo, Celotti \& Fabian (1997b, hereafter DCF97) have considered a model
in which energy is released via magnetic field reconnection in localized active
regions forming a patchy corona above an optically-thick accretion disc. They
assume that the energy density of the magnetic field is in equipartition with
that of the local radiation,
\begin{equation}
{B^2\over 8\upi}\approx {9 \over 16\upi} {L\over N (r_{\rm b} R_{\rm g})^2 c},
\label{e:radfield}
\end{equation}
where $r_{\rm b} R_{\rm g}$ is the radius of an active blob and $N$ is the
average number of blobs. The right-hand-side of this equation corresponds to
the average photon density in an optically thin, uniformly-radiating, sphere.
This yields values of $B$ somewhat larger than those of DCF97, whose used a
numerical factor of $1/4\upi$ in their expression for $B$. Consequently, our
estimates of $L_{\rm CS}$ will be slightly higher than those obtained using the
formalism of DCF97. In the case of an optically and geometrically thin
reconnection region, the numerical factor above should be $9/12\upi$. Equation
\odn{e:radfield} can be rewritten as
\begin{equation}\label{B2}
B\approx 10^9 {1 \over r_{\rm b}} \left(\frac{\eta}{N m}\right)^{1/2} {\rm G}.
\end{equation}
Following DCF97, we will assume hereafter $N= 10$. The characteristic size of a
reconnection region has been estimated by Galeev et al.\ (1979) as
\begin{equation}
\label{e:height}
r_{\rm b}  \sim h_{d} \alpha_{\rm v}^{-1/3},
\end{equation}
where $h_{\rm d} R_{\rm g}$ is the scale height of an underlying
optically-thick disc, and $\alpha_{\rm v}$ is the standard viscosity parameter.
This typically yields a range of $0.01 \la r_{\rm b} \la 5$, depending on the
disc parameters and increasing with accretion rate (see,e.g.\ Svensson \&
Zdziarski 1994). On the other hand, the specific corona model of Galeev et al.\
(1979) can power only very weak coronae (Beloborodov 1999) and other mechanisms
can be responsible for formation of luminous coronae. Thus, we simply assume
$r_{\rm b} =2$ in numerical examples below (which value was also used by
DCF97).

With the above value of $B$, we can derive the ratio of the CS luminosity from
the corona to the total coronal luminosity,
\begin{equation}\label{ratio2}
{L_{\rm CS}\over L}\approx
\frac{650 \varphi (\eta/N m r_{\rm b}^2 )^{0.46\alpha-0.09}\Theta^{3.78-0.11
\alpha} }{10^{2.24 \alpha} \alpha^{0.13+0.06 \alpha}(1-\alpha)}
\end{equation}
for $0.4\la \alpha\la 0.9$. The analogous formula for $\alpha\ga 1.1$ is
\begin{equation}\label{soft2}
{L_{\rm CS}\over L}\approx 1.25
\left(1+\frac{3\varphi}{\alpha-1}\right)
\left(\eta\over N m r_{\rm b}^2\right)^{0.37}{\Theta^{3.66}\over
\alpha^{0.19}}.
\end{equation}

From equations (\ref{ratio2})-(\ref{soft2}), we find that, in most cases, the
coronal model yields much lower values of $L_{\rm CS}$ than the model of a hot
accretion flow of Section \ref{hot_ions}. For the hard state of Cyg X-1 with
$\eta=0.02$ and $m=10$ (Section \ref{hot_ions}), and for $N=10$, $r_{\rm b}=2$,
we find $L_{\rm CS}/ L \approx 1.7\times 10^{-2}$, which is $\sim 20$ times
less than $L_{\rm CS}/L$ obtained in the hot flow model. Only an extremely low
and inconsistent with equation (\ref{e:height}) value of $r_{\rm b}\sim
10^{-5}$ would lead to $L_{\rm CS}=L$ in this case. For objects with softer
spectra, even lower values of $L_{\rm CS}/ L$ are obtained. For objects with
low luminosities, Comptonized bremsstrahlung would dominate, see equation
(\ref{lcb}), which applies to all coronal models discussed here with $r=N
r_{\rm b}$.

Thus, thermal synchrotron radiation cannot be a substantial source of seed
photons for Comptonization in active coronal regions with equipartition between
magnetic field and radiation, even for weak sources.  This finding can be
explained by considering consequences of this equipartition in the presence of
strong synchrotron self-absorption.  With neither self-absorption nor
additional sources of seed photons, equipartition between the energy densities
in the field and in photons leads to the net Comptonization luminosity, $L_{\rm
C}$, roughly equal to the synchrotron luminosity, $L_{\rm S}$, which then
implies $\alpha\sim 1$.  (An additional assumption here is the validity of the
Thomson limit, e.g.\ Rybicki \& Lightman 1979.)  However, strong
self-absorption in a thermal plasma with parameters relevant to compact sources
dramatically reduces $L_{\rm S}$, i.e., $L_{\rm S}\ll L_{\rm C}$.  Then, a very
hard Comptonization spectrum, with $\alpha\sim 0$, is required to upscatter the
few synchrotron photons into a spectrum with luminosity of $L_{\rm C}$.  On the
other hand, if $\alpha\sim 1$, as typical for compact cosmic sources, CS
photons would give only a tiny contribution to the actual coronal luminosity,
$L$, and other processes, e.g.\ Comptonization of blackbody photons from the
underlying disc would have to dominate.

This conclusion differs from findings in DCF97 as well as in Di Matteo, Celotti
\& Fabian (1999) that this process is often important under conditions typical
to compact objects, in particular in GX 339--4.  This discrepancy stems mostly
from their assumption of $\eta=1$ (i.e., $L = L_{\rm E}$) for calculating the
value of $B$ in those papers (T.~Di Matteo, private communication). In the
specific cases they consider, $L_{\rm CS}\ll L_{\rm E}$, i.e., only a small
fraction of the dissipated power is radiated away, and they postulate that the
remaining power is stored in magnetic fields (see Section 2.2 in DCF97).
However, they do not consider the final fate of the power supplied to the
magnetic fields.  In fact, the supplied power has to be eventually either
radiated away or transported away from the disc.  In the former case, we would
recover our result of $L_{\rm CS}\ll L$. In the latter, the power in magnetic
fields could be either converted to kinetic power of a strong outflow or
advected to the black hole.  Those possibilities require studies that are
beyond the scope of this paper.  We only note that a physical realization of a
strong but effectively non-radiating outflow in the vicinity of a luminous
binary appears difficult.  Similarly, the advection time scale of an
optically-thick disc is probably much longer than the time scale for
dissipation of coronal magnetic fields.  Finally, we note that $\eta=1$ would
often require accretion rates much higher than those estimated from the
physical parameters of X-ray binaries (see, e.g., Zdziarski et al.\ 1998 for
the case of GX 339--4).

In addition, the values of $\Theta$ assumed in DCF97 and Di Matteo et al.\
(1999) for luminous states of black-hole sources are significantly higher than
those used in the examples shown in this work (which were derived from the best
currently-available data). Due to the strong dependence of $L_{\rm CS}$ on
$\Theta$ [see eqs.\ (\ref{ratio2})-(\ref{soft2})], this also results in higher
values of $L_{\rm CS}$.

We note that equation (\ref{ratio2}) can yield $L_{\rm CS}\sim L$ for
$\Theta\ga 1$. This is partly explained by most of energy density in
Comptonized photons being then in the Klein-Nishina limit. Then, the energy
density of photons in the Thomson limit is much less than the total energy
density, and thus more comparable with energy density of self-absorbed
synchrotron photons. Thus, the CS process is expected to play an important role
in sources with $\Theta\ga 1$. Note, however, that equations
(\ref{ratio2})-(\ref{soft2}) break then down as the approximations of equations
\odn{nu_ratio} and \odn{alpha} become invalid and numerical calculations should
be employed in that case.

\subsubsection{Dissipation of magnetic field}
\label{dissipation}

As discussed by, e.g., Di Matteo, Blackman \& Fabian (1997a), the strength of
the coronal magnetic field is expected to be higher than the equipartition
value of equation (\ref{e:radfield}). If an active region is powered by
dissipation of magnetic field and the dissipation velocity is $g v_{\rm A}$,
where $g\leq 1$ and $v_{\rm A} = B/(4\upi n_{\rm e} m_{\rm p} \mu_{\rm
e}/\mu_{\rm i})^{1/2}$ is the Alfven speed, the energy density stored in the
field will be $c/g v_{\rm A}$ times that in radiation (which escapes the source
with the velocity $c$), and $h_{\rm b}/r_{\rm b} \sim g$, where $h_{\rm b}
R_{\rm g}$ is the scale height of the active region (Longair 1992). Then, for a
given $g=h_{\rm b}/r_{\rm b}$ and accounting for the dependence of the
radiation
density on geometry, we have,
\begin{equation}
{B^2\over 8\pi}\frac{g v_{\rm A}}{c}\approx
{9 \over 4(3+g)\upi } {L \over N (r_{\rm b} R_{\rm g})^2 c}.
\label{e:radfield2}
\end{equation}
Using the approximation of equation \odn{alpha}, we have then
\begin{equation}
\label{disf}
B\approx
\frac{9.8\times10^8 \eta^{1/3}}{r_{\rm b}^{5/6}
m^{1/2}\alpha^{5/24}\Theta^{5/24} [g(3+g) N]^{1/3} }\,
{\rm G}.
\end{equation}

The luminosity in the CS emission is then approximately given by
\begin{eqnarray}\label{ratio2a}
\lefteqn{ {L_{\rm CS}\over L}\approx \frac{312\,(1+g)}{10^{2.25 \alpha}
[g(3+g)]^{0.61+0.3\alpha}}\times} \nonumber \\
\lefteqn{ \quad \frac{\Theta^{3.4-0.3\alpha}r_{\rm b}^{0.48-0.76\alpha}
(N/\eta)^{0.39-0.3\alpha}
\varphi} { m^{0.46\alpha-0.09}
\alpha^{0.5+0.25\alpha}(1-\alpha)}, }
\end{eqnarray}
for $0.4\la \alpha\la 0.9$, and
\begin{equation}\label{ratio2b}
{L_{\rm CS}\over L}\approx
0.59 \left(1+\frac{3 \varphi}{\alpha-1}\right)
\frac{(1+g)\Theta^{3.09}(N/\eta)^{0.09}}{ [g(3+g)]^{0.91} m^{0.37} r_{\rm
b}^{0.28}\alpha^{0.76} },
\end{equation}
for $\alpha \ga 1.1$, where we accounted for the dependence of the source area
on $g$.

We note that Di Matteo et al.\ (1997a) adopted $g = h_{\rm b}/r_{\rm b} = 0.1$,
whereas Di Matteo (1998) assumed $g=1$ but $h_{\rm b}/r_{\rm b} = 0.1$. We set
$g=h_{\rm b}/r_{\rm b}=0.1$ in our examples below.

The values of $L_{\rm CS}$ in this model are much higher than those in the
previous case, but still relatively low. For Cyg X-1 for the same parameters as
above ($\eta=0.02$, $N=10$, $r_{\rm b}=2$), we find $L_{\rm CS}/L \sim 1$, i.e.
this model can, in principle, account for the luminosity of this object.
However, for objects with softer spectra, we obtain $L_{\rm CS}/L\ll 1$, as
shown in Figure \ref{comparison}b (which shows comparison with the same data as
Figure \ref{comparison}a, and $\eta=0.01$ is assumed for the model curve
corresponding to the average Seyfert-1 spectrum).  We see that the predictions
of this model for luminous sources, with $\eta\ga 0.01$, are qualitatively
similar to those of the hot-flow model.

On the other hand, this model predicts values of $L_{\rm CS}$ for weak objects
much lower than those in the hot disc model.  This can be seen by comparing the
dependence on $\eta$ in both cases.  In the case of a hot flow, $L_{\rm CS}$
does not depend on $L$, so formally $L_{\rm CS}/L \propto \eta^{-1}$ (not
including dependences of the plasma parameters on $L$). On the other hand,
equations \odn{ratio2a}-\odn{ratio2b} yield roughly $L_{\rm CS}/L \propto
\eta^{-0.1}$ in the present case (which difference is mainly caused by the
source area much smaller in the patchy corona model than in the hot flow
model).  Thus, decreasing $L$ leads to only a marginal increase of $L_{\rm
CS}/L$.  In contrast, the luminosity from Comptonized bremsstrahlung is about
the same in the two models.  Consequently, $L_{\rm CB}\gg L_{\rm CS}$ is
typical of weak sources in the coronal models.

This is indeed the case for NGC 4258, where we have found it impossible to
reproduce its 2--10 keV power-law index and luminosity even with $r_{\rm b} \ll
1$, except for $\Theta > 1$. Although such a high temperature cannot be ruled
out observationally at present, the corresponding $\ts\ll 1$, which then would
lead to a curvy spectrum reflecting individual scattering profiles. Then, the
similarity of the X-ray spectral index of that object to the average index of
Seyfert 1s would have to be accidental, which we consider unlikely.

\begin{figure}
\begin{center} \noindent
\epsfxsize=7.2cm \epsfbox{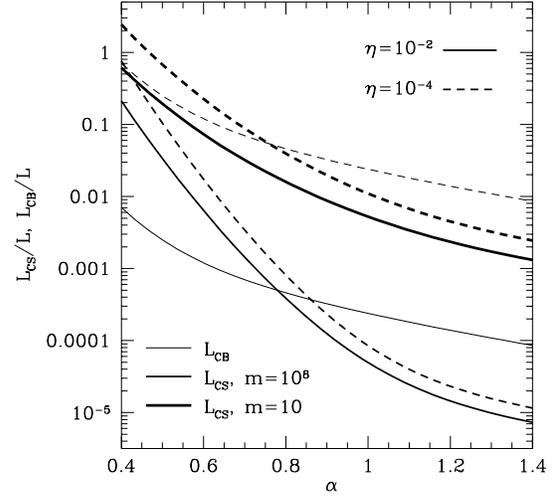}\end{center}\nobreak
\caption{The fraction of the total coronal luminosity produced by the CS
process (heavy and medium curves for $m=10$ and $m=10^8$, respectively) and by
Comptonized bremsstrahlung (thin curves applying to both values of $m$) as a
function of the spectral index of the CS radiation, for $\eta=10^{-2}$ (solid
curves) and $\eta=10^{-4}$ (dashed curves) in the case of magnetic field of
equation (\ref{e:radfield2}). See Section \ref{dissipation} for other
parameters.}
\label{brem_eff}
\end{figure}

Figure \ref{brem_eff} shows a comparison of CS and Comptonized bremsstrahlung
emission as functions of $\alpha$ for two values of $m$ and two values of
$\eta$, and for $g=0.1$, $N=10$, $r_{\rm b}=2$, $\Theta=0.2$.  As expected,
bremsstrahlung dominates at low values of $\eta$ and high values of $\alpha$,
and its Eddington ratio is independent of $m$.

\subsubsection{Magnetic field in the disc}
\label{disc}

The magnetic field strength in a corona is limited by equation 
\odn{e:radfield2} provided most of the energy of the field is dissipated there, 
which seems to be a likely assumption.  However, that field depends on the 
factor $g$, whose value appears uncertain, and which could, in principle, be 
$\ll 1$.  Furthermore, we cannot rule out the presence of a stronger, more 
permanent field, which would dissipate only partially.  In any case, the 
coronal magnetic field strength will be lower than that inside the disc, 
$B_{\rm d}$.  The latter is in turn limited by equipartition with the disc 
pressure, as upward buoyancy forces will rapidly remove any excess magnetic 
flux tubes from the disc (Galeev et al.\ 1979).  Such an equipartition field 
was adopted by Di Matteo (1998) in calculations of the rate at which energy was 
supplied to a corona.  On the other hand, if the viscosity is provided alone by 
the magnetic field, its pressure was estimated as $\alpha_{\rm v}$ times the 
disc pressure by Shakura \& Sunyaev (1973), which would approximately 
correspond to multiplying equations 
(\ref{e:disc_gasfield})-(\ref{e:disc_radfield}) below by $\alpha_{\rm 
v}^{1/2}$. Numerical simulations of Hawley, Gammie \& Balbus (1995; see Balbus 
\& Hawley 1998 for a review) also confirm that the magnetic field in the disc 
always remains below the equipartition value.

Here, we utilize formulae for the vertically-averaged disc structure of
Svensson \& Zdziarski (1994), which take into account that a fraction, $f<1$,
of the energy dissipated in the disc is transported to the corona (see a
discussion on the transport mechanisms in Beloborodov 1999). However, we
recalculate the disc structure taking into account the additional pressure
provided by the equipartition magnetic field. For consistency with the rest of
this work, we assume equipartition with energy density rather than with
pressure, which assumption has only a minor effect on our results. Expressions
for $B_{\rm d}$ below are computed at $r=11$, which is approximately the radius
at which a radiation-dominated region first appears with increasing accretion
rate (Svensson \& Zdziarski 1994). For lower radii, somewhat higher values of
$B_{\rm d}$ are obtained, but the contribution of those radii to the total
dissipated power is small. In a disc region dominated by the gas pressure, we
have then,
\begin{equation} \label{e:disc_gasfield} B_{\rm d} \approx \frac{1.1
\times 10^{8} \dot{m}^{2/5}}{(\alpha_{\rm v} m)^{9/20} (1-f)^{1/20}}\, {\rm G},
\end{equation}
where $\dot{m}$ is the dimensionless accretion rate, $\dot m \equiv \dot{M}
c^2/L_{\rm E}$. In a disc region dominated by radiation pressure, we have,
\begin{equation} \label{e:disc_radfield} B_{\rm d} \approx
5.2 \times 10^{7} \left[\alpha_{\rm v} m (1-f) \right]^{-1/2}\ {\rm G}.
\end{equation}
Note that the radiation-pressure dominated region appears only for accretion
rates higher than certain value $\dot{m}_{\rm crit}$. Hereafter, we will assume
$\alpha_{\rm v}=0.1$ and that half of the accreted energy is dissipated in the
corona, i.e., $f=1/2$. We assume the efficiency of cooling-dominated accretion
in the Schwarzschild metric, which gives the power dissipated in the corona of
$L=0.057 \dot{m} f L_{\rm E}$.

We first compare the maximum strength of coronal magnetic field of equation
\odn{e:radfield2} with the disc field of equations
\odn{e:disc_gasfield}-\odn{e:disc_radfield}. For parameters relevant to compact
objects, the disc field is always larger than the maximum coronal magnetic
field, up to two orders of magnitude, except for extremely low accretion rates,
$\dot{m} \la 10^{-10}$. This shows the self-consistency of the coronal models
of Sections \ref{radiation}-\ref{dissipation}.

As discussed above, given the uncertainties about the mechanism of
magnetic field reconnection in active regions above accretion discs, we cannot
rule out the presence of an average coronal field strength, $\langle B\rangle$,
stronger than that of equation \odn{e:radfield2}, and limited by equations
\odn{e:disc_gasfield}-\odn{e:disc_radfield},
\begin{equation}\label{Bd}
\langle B\rangle =\varepsilon B_{\rm d},
\end{equation}
where $\varepsilon<1$ accounts for an inevitable decay of the field during a
flare event. The dissipation rate of the magnetic field in a
reconnection event is $\propto B^2$ (see e.g.\ section 12.4 in Longair 1992) so
the field decays exponentially. Since the above average is weighted by the
luminosity of an active region, which quickly decreases with decreasing $B$,
$\varepsilon$ can be, in principle, of the order of unity. Thus, we adopt
$\varepsilon=0.5$ in numerical examples below.

Assuming the magnetic field strength given by equation (\ref{Bd}), we obtain
for disc regions dominated by the gas pressure,
\begin{eqnarray}
\lefteqn{\frac{L_{\rm CS}}{L} \approx
\frac{200 \times 10^{-3.1 \alpha} }{\alpha^{0.13+0.06\alpha}(1-\alpha)}\times}
\nonumber\\
\lefteqn{\qquad \frac{\varepsilon^{1.82+0.91\alpha}N r_{\rm b}^2
\Theta^{3.78-0.11\alpha}
\dot{m}^{0.36\alpha-0.27}\varphi}{\alpha_{\rm v}^{0.82+0.41 \alpha} f
(1-f)^{0.09+0.05\alpha} m^{0.41\alpha-0.18} }}
\end{eqnarray}
for $0.4 \la \alpha \la 0.9$ and
\begin{equation}
\frac{L_{\rm CS}}{L} \approx 0.052
\left(1+\frac{3 \varphi}{\alpha-1}\right)
\frac{\varepsilon^{2.73}N r_{\rm b}^2 \Theta^{3.66}
\dot{m}^{0.09}}{\alpha_{\rm v}^{1.23} f (1-f)^{0.14} m^{0.23}
\alpha^{0.19}}
\end{equation}
for $\alpha \ga 1.1$. Similar expressions can be derived for regions dominated
by the radiation pressure.

\begin{figure}
\begin{center} \noindent
\epsfxsize=7.2cm \epsfbox{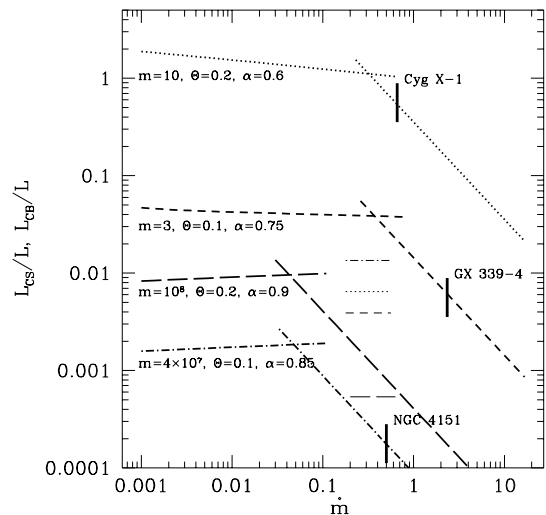}\end{center}\nobreak
\caption{The ratio of the CS and bremsstrahlung luminosities (heavy and thin
lines, respectively; the latter ratio is independent of $\dot m$) to the power
dissipated in a patchy corona as functions of the accretion rate in the case of
the coronal magnetic field strength equal to 0.5 of that inside the disc. The
solutions corresponding to dominance of the gas pressure and radiation pressure
are shown for $\dot{m}$ below and above $\sim \dot{m}_{\rm crit}$,
respectively. The chosen values of $m$, $\Theta$, $\alpha$ are the same
as in Figure \ref{comparison}. See Section \ref{disc} for details.
The vertical bars mark our estimates of the accretion rates in 3 cases.}
\label{comparison2}
\end{figure}

Figure \ref{comparison2} shows the above ratios for the same values of
$\alpha$, $\Theta$ and $m$ (except that $m=10^8$ was assumed for the average
Seyfert 1 spectrum) as in Figure \ref{comparison} as functions of $\dot{m}$ for
$\varepsilon=0.5$, $r_{\rm b}=2$, and $N=10$.  We see that the obtained
luminosity is comparable to the one dissipated in the corona only in the case
of Cyg X-1.  In other cases, the CS process gives a small contribution,
comparable to or lower than the contribution of bremsstrahlung.

In the case of NGC 4258, we can reproduce the 2--10 keV power-law index and
luminosity with $\Theta=0.43$ and $r_{\rm b}=2$.  We note, however, that the CS
luminosity is a sensitive function of the size of active regions in this model,
$L_{\rm CS} \propto r_{\rm b}^2$ (which dependence is much stronger than that
in the previous coronal models).  Based on equation (\ref{e:height}), we expect
$r_{\rm b}$ decreasing with the decreasing $L$.  If this is indeed the case for
NGC 4258, the model of coronal CS emission could be then ruled out for this
object (unless $\Theta> 1$, as in the case discussed in Section
\ref{dissipation} above).

We therefore again conclude that, in coronal models, the CS radiation can be 
important only in the case of stellar-mass sources with hard spectra. This 
radiative process is negligible in the case of objects with soft spectra 
(probably including low-luminosity sources) when Comptonization of photons from 
a cold accretion flow is expected to dominate. Note that if the disc magnetic 
field pressure is less than that of equipartition, e.g., by $\alpha_{\rm v}$ 
(Shakura \& Sunyaev 1973), the predicted upper limits of $L_{\rm CS}$ would be 
even lower (e.g.\ by an order of magnitude for $\alpha_{\rm v}=0.1$).

We can compare our conclusions with those of Ghisellini, Haardt \& Svensson
(1998), who also consider magnetically dominated, patchy coronae and compare
the relative importance of Comptonization of synchrotron and disc photons.
They, however, do not assume a thermal electron distribution but instead assume
a given form of electron acceleration and then calculate the electron
distribution taking into account a balance between the acceleration, escape,
Compton cooling and synchrotron emission and reabsorption. A resulting electron
distribution consists typically of a quasi-thermal hump and a high-energy tail,
which is an important difference with respect to the calculations presented
here. Furthermore, they assume a constant magnetic field for a series of models
with varying luminosity, and their magnetic field is not directly compared to
that produced in the disc.

Still, their findings are compatible with ours. Namely, among the specific
models they present, the CS process dominates only when $\Theta \sim 1$ and the
magnetic energy density is more than one order of magnitude higher than that of
radiation, see their Figure 4. We find that the parameters of their models with
the CS process dominating, correspond approximately to $\eta \sim 10^{-4}$,
i.e.\ to low-luminosity sources, in our model above (assuming $r_{\rm b} \sim
1$ and $N\sim 10$). Then, at $\Theta \sim 1$, we obtain predictions about
the significance of the CS process similar to theirs.

\section{Applications to accreting neutron stars} \label{neutron}

Power-law X-ray spectra are, in fact, observed not only from black-hole
sources.  Similar power laws are detected, e.g., from binary systems with
weakly-magnetized neutron stars in the so-called low spectral state.
Unambiguous classification of some of those sources as neutron stars comes from
detections of type-1 X-ray bursts from them.  Those spectra are often modelled
by thermal Comptonization (e.g.\ Barret et al.\ 1999, hereafter B99), in which
case the question arises of the origin of seed photons.

It is also of interest that accreting weakly-magnetized neutron stars show a
number of similarities to accreting black holes (see e.g.\ a discussion in
B99).  First, this concerns timing properties in X-rays, namely time lags
between soft and hard X-rays, which suggests similar geometries of the sources
and/or similar variability mechanisms (e.g.\ Ford et al.\ 1999).  Second,
spectra of black holes in their hard states fit remarkably well a correlation
between the 1--20 and 20--200 keV luminosities seen in neutron-star binaries
(Barret, McClintock \& Grindlay 1996) which may indicate that similar radiation
mechanisms are operating in the two classes of objects. On the other hand, the
presence of the neutron star results in more sources of soft photons than in
the case of a black hole, namely thermal radiation from the surface of the star
and synchrotron radiation in the stellar magnetic field.

Cyclotron radiation as a source of soft photons Comptonized in weakly
magnetized accreting neutron stars has recently been considered by Psaltis
(1998). In his model, self-absorbed cyclotron photons are produced in a layer
above the stellar surface with $kT \sim 20$ keV and a magnetic field of $B \la
10^{10}$ G, and then subsequently Comptonized in a hot spherical corona of $kT
\sim 10$ keV. Here, we consider a more generic geometry in which synchrotron
photons are produced and Comptonized in the same medium, the temperature of
which is determined from observations. This leads us to investigating plasmas
of temperatures higher than those considered by Psaltis (1998). As in Section
4, we compare the luminosities in our model spectra to those observed.

We analyze here spectra of four X-ray bursters. The first one is 4U~0614+091,
which low-state spectrum was fitted with the Comptonization model of Poutanen
\& Svensson (1996) by Piraino et al.\ (1999). They found the seed blackbody
photons have $kT_{\rm BB}\la 30$ eV, and the plasma parameters are given by
$\Theta\simeq 0.5$ and $\alpha=1.44$. The luminosity was $L_{\rm
1-200keV}\simeq 3.7 \times 10^{36}$ erg~s$^{-1}$ assuming $D=3$ kpc (Brandt et
al.\ 1992). It is interesting that the above electron temperature is much
higher than that seen from other X-ray bursters so far. The parameters of the
second object in our sample, 1E~1724--3045, modeled by B99 with a
Comptonization model of Titarchuk (1994) are $kT_{\rm BB}\simeq 1$ keV,
$\Theta\simeq 0.057$, $\alpha=0.95$. (Since B99 did not give the spectral index
of their fit, we have obtained it from the Comptonization model they used.) The
luminosity, given $D=6.6$ kpc (Barbuy, Bica \& Ortolani 1998), is $L_{\rm
1-200keV}\simeq 1.3 \times 10^{36}$ erg~s$^{-1}$. The third source is
GS~1826--238, for which B99 found $\alpha=0.73$, and $\Theta\simeq 0.08$ from
the high-energy tail with the model of Poutanen \& Svensson (1996). At $D=7$
kpc (B99), $L_{\rm 1-200keV}\simeq 1.5 \times 10^{36}$ erg~s$^{-1}$.
Observations of the fourth object, XB~1916--053, were reported by Church et
al.\ (1998). The X-ray spectrum was fitted as a power-law of index
$\alpha=0.61$ and the temperature was estimated from the cut-off in the
spectrum to be $\Theta\simeq 0.06$. At $D=9$ kpc, $L_{\rm 0.5-200keV}\simeq 1.1
\times 10^{36}$ erg s$^{-1}$ (Church et al.\ 1998).

We consider then a generic model with a spherical, uniform, plasma cloud, and
investigate the relation between the size of the cloud and the magnetic field
strength required to reproduce the observed luminosities at their spectral
indices and temperatures. From equations \odn{nu_ratio} and \odn{softL}, we
expect that this relation will be roughly of the form of $B \propto R^{-0.7}$.
We note that since a CS spectrum is normalized by its self-absorbed,
optically-thick, part, the luminosity is roughly proportional to the source
area, with the source geometry being only of secondary importance. Thus, our
results may also approximate those for emission from a thin layer above the
stellar surface. A second constraint on the model can be obtained from the
fitted temperature of the seed photons, which is known for 4U~0614+091 and
1E~1724--3045. In those cases, we can identify the seed-photon temperature with
the maximum possible turnover energy, $h \nu_{\rm t} \la kT_{\rm BB}$, which
then yields an upper limit on $B$. This limit is $B \la 10^8$ G and $B \la 6
\times 10^9$ G, respectively. In the case of GS~1826--238, the seed-photon
temperature was not fitted, but we can estimate, from the shape of the spectrum
presented by B99, that $kT_{\rm BB}\la 1$ keV, which yields $B \la 5 \times
10^9$ G. We have no corresponding limit for XB~1916-053.

\begin{figure}
\begin{center} \noindent
\epsfxsize=7.2cm \epsfbox{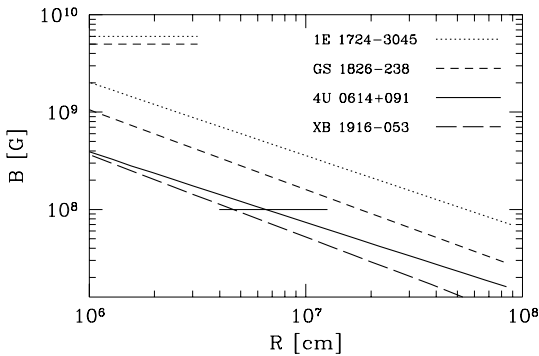}\end{center}\nobreak
\caption{
The relation between the size of the emission region and its magnetic field
strength required for the CS process to reproduce the observed  X-ray spectra
and luminosities of four X-ray bursters (heavy lines). The thin horizontal
lines show the upper limits on $B$ resulting from fitted constraints on the
energy of seed photons. Both constraints have to be satisfied for the CS
process to dominate.}
\label{xrb}
\end{figure}

The size--magnetic field relation obtained for the four objects is shown in
Figure \ref{xrb}. Those results suggest that Comptonization may take place in a
corona of the size comparable to the stellar radius ($\sim 10^6$ cm). The
required magnetic field is then $B \la 10^9$ G. This value is consistent both
with the upper limits resulting from fitting the energy of seed photons and
with the maximum field strength possible in X-ray bursters, $B \la
10^{10}$--$10^{11}$ G, see Lewin, van Paradijs \& Taam (1995). (Their
constraint results from the requirement that the magnetic field does not funnel
matter towards the magnetic poles, and more stringent constraints can be
obtained for a specific structure of the accretion flow.)

We note that if Comptonization takes place close to the stellar surface,
reflection albedo close to unity would be required. Otherwise reprocessing in
the surface layers would result in a strong blackbody component, which is not
seen.

We also find that it is unlikely that the size of the region where synchrotron
photons (dominating the supply of seed photons) are produced and Comptonized is
much larger than the neutron star radius. Such a configuration would be then
similar to that of an optically thin disc around a black hole (Section
\ref{hot_ions}) or an advection-dominated extended corona above a cold disc
(Narayan, Barret \& McClintock 1997). Then the required magnetic field, $B \sim
10^8$ G at $R \sim 10^7$ cm, would be too strong to be sustained by the disc
[see equation \odn{B}, where we estimate the magnetic field in hot, optically
thin, discs].

\section{Discussion}
\label{discussion}

A general feature of models of accretion flows around black holes considered
above is the dependence of $B\propto m^{-1/2}$ [or very close to it, equation
(\ref{e:disc_gasfield})]. This then implies decreasing values of $L_{\rm
CS}/L_{\rm E}$ with $m$, and the relative importance of the CS process much
weaker in AGNs than in black-hole binaries. The predicted values of $L_{\rm
CS}$ strongly increase with increasing spectral hardness and temperature of the
plasma. Compared with observations, we find that both the hot flow and coronal
models can, in principle, explain emission of the hardest ($\alpha\sim 0.5$)
among luminous (Eddington ratios of $\sim 0.01$), stellar-mass, sources in
terms of the CS process. However, this process provides a negligible
contribution to the luminosity, $L$, of luminous stellar-mass sources with soft
spectra and of luminous AGNs regardless of $\alpha$. Taking into account the
overall similarity between properties of luminous sources with either hard or
soft spectra and containing either stellar-mass or supermassive black holes
(e.g., Zdziarski 1999), it is likely that the CS process is in general
energetically negligible in luminous sources.

However, the hot flow and coronal models differ in their predictions for the
relative importance of the CS process with decreasing $L/L_{\rm E}$. In the hot
flow model, there is clearly a range of $L/L_{\rm E}\ll 0.01$, in which the CS
process can yield a dominant contribution to $L$ even in the case of AGNs. On
the other hand, the coronal models have rather weak dependences of $L_{\rm
CS}/L$ on $L/L_{\rm E}$. However, at some low value of $L/L_{\rm E}$,
bremsstrahlung emission becomes dominant. Therefore, we have found that, in the
coronal models, there is no range of $L/L_{\rm E}$ (except for hard,
stellar-mass, sources, see above) in which the CS process could dominate
energetically. In both models, bremsstrahlung is expected to give the main
contribution to $L$ in very weak sources.

There are three complications in the picture above. First, the plasma
parameters are expected, in general, to depend on $L/L_{\rm E}$, which will
modify theoretical dependences of $L_{\rm CS}/L$ on the Eddington ratio. This
is especially the case of the electron temperature, which is expected to be
higher in low-luminosity sources, due to less efficient plasma cooling. This
would strongly increase the efficiency of the CS process. We did not consider
temperatures much higher than those observed in luminous objects, thus our
conclusions about the role of the CS process in low-luminosity sources are
rather conservative. Unfortunately, we have as yet no observational constraints
on the plasma temperature in such sources.

Second, we have considered here accretion onto non-rotating black holes.
Black-hole rotation increases, in general, the efficiency of accretion, and the
CS process may yield then luminosities higher than those found here.  This is,
in fact, confirmed by a study of Kurpiewski \& Jaroszy\'nski (1999), who have
considered advection-dominated flows and have found that an increase of the
black-hole angular momentum leads to increase of the CS emission.  This happens
due to both the synchrotron emission itself as well its Comptonization becoming
much more effective in innermost regions of the disc.  Analysis of those issues
is, however, beyond the scope of this work. Third, even when other processes
dominate energetically, there can be a range of photon energies in which the CS
process gives the dominant contribution.

Still, the strong conclusion of Section \ref{applications} is that the thermal
CS process does not dominate observed X-ray spectra of luminous black-hole
sources.  Rather, the dominant X-ray producing process in those sources appears
to be Comptonization of blackbody photons emitted by some optically-thick
medium, e.g., an optically-thick accretion disc or clumps of cold matter.
Observationally, the presence of such component is indicated by the wide-spread
presence of Compton reflection from a cold medium in luminous black-hole
sources (e.g.\ Zdziarski et al.\ 1999) as well as observations of
blackbody-like soft X-ray components in spectra of black-hole binaries (e.g.\
Ebisawa et al.\ 1994).  Furthermore, a very strong correlation between the
reflection strength and the X-ray spectral index is present in those sources
(Zdziarski et al.\ 1999), in particular in Cyg X-1 and GX 339--4 (Gilfanov,
Churazov \& Revnivtsev 1999; Revnivtsev, Gilfanov \& Churazov 1999).  The
presence of such correlation is naturally explained in models with plasma
cooling due to blackbody emission of cold matter, whereas it cannot be
explained if the CS process dominates.

On the other hand, we find in Section \ref{neutron} that the thermal CS process
can easily account for power law emission of weakly magnetized neutron stars in
their low states. A possible caveat for that model is that a correlation
between reflection and $\alpha$ has been observed in 4U~0614+091 (Piraino et
al.\ 1999), although its interpretation remains ambiguous due to the lack of a
detection of blackbody emission strong enough to provide seed photons for
Comptonization.

We point out that a small non-thermal tail in the electron distribution due to
acceleration (as in the model of Ghisellini et al.\ 1998), e.g.\ stochastic or
in reconnection events could significantly increase the turnover frequency
(since electrons with $\gamma \gg 1$ are usually responsible for emission at
the turnover frequency), and thus increase the CS luminosity from accreting
black holes and neutron stars. This issue is the subject of our work in
progress.

Finally, we notice thermal Comptonization may be (at most) a secondary process
in spectral formation in some classes of compact X-ray sources. One important
class of such sources are black-hole binaries in the soft state, which X-ray
and soft \g-ray spectra appear non-thermal (e.g.\ Gierli\'nski et al.\ 1999).
In that case, the X-ray spectrum contains a very strong blackbody component,
which probably dominates over synchrotron photons as seeds for non-thermal
Comptonization.

\section{Conclusions}

The main results of this work can be outlined as follows.

We have derived and tested analytic approximations for the synchrotron
emission coefficient in thermal plasmas, applicable especially to the
semi-relativistic range of temperatures, $T \sim 10^9$ K. We have also obtained
analytic approximations to the turnover frequency (at which the plasma becomes
optically-thick to absorption).

Then, we have presented a method to treat thermal Comptonization of synchrotron
radiation. Our approximate analytic expressions allow for self-consistent
calculations of the spectrum and the luminosity of such a source and can be
easily applied to models of accretion discs.

We have also investigated the role of the thermal CS process in accretion flows
around stellar-mass and supermassive black holes. We have considered two main
scenarios: a hot, two-temperature, optically-thin flow and active regions above
a cold, optically-thick disc. We have found that this process is only
marginally important in luminous X-ray sources containing accreting black
holes, and it can possibly dominate only in stellar-mass sources with hardest
spectra. The dominant radiative process in those sources appears to be
Comptonization of blackbody radiation emitted by cold matter in the vicinity of
the hot plasma. On the other hand, the CS process can explain X-ray spectra of
weaker sources, e.g., low-luminosity AGNs, but only in the hot-flow model.
Finally, below certain low luminosity, bremsstrahlung becomes the dominant
process.

Finally, we considered the case of weakly-magnetized accreting neutron stars.
We have found that their power-law X-ray spectra in the low state can be
accounted for by the CS process taking place in a corona of the size comparable
to the stellar radius.

\section*{ACKNOWLEDGMENTS}

This research has been supported in part by a grant from the Foundation for
Polish Science and the KBN grants 2P03D00624 and 2P03D01716.  It is a pleasure
to acknowledge valuable discussions with Chris Done, Marek Gierli\'nski,
Krzysztof Jahn and Marek Sikora. We are grateful to Andrei Beloborodov, Vahe
Petrosian, and Tiziana Di Mateo, the referee, for thoughtful comments on this
work.

\bsp

\label{lastpage}

\end{document}